\documentclass[%
 reprint,
 amsmath,amssymb,
 aps,
 prl,
]{revtex4-2}

\usepackage{newtxtext}
\usepackage[smallerops]{newtxmath}

\usepackage{graphicx}
\usepackage{dcolumn}

\usepackage{bm}
\usepackage{mathtools}
\usepackage{xcolor}

\usepackage{tikz}
\usetikzlibrary{calc}

\usepackage{hyperref}

\def\ang#1{\left\langle #1\right\rangle}

\newcommand{\iu}{\mathrm{i}\mkern1mu}

\def\hmf{{\tilde h}}

\newcommand{\PRLsection}[1]{\emph{#1}---}

\DeclareMathOperator{\Tr}{Tr}

\begin{document}

\preprint{APS/123-QED}

\title{Storing Infinite Dynamical Attractors in Nonreciprocal Associative Neural Networks}

\author{Miguel Aguilera}
\email[Corresponding author: ]{maguilera@bcamath.org}
\affiliation{BCAM -- Basque Center for Applied Mathematics, 48009 Bilbao, Spain}
\affiliation{IKERBASQUE, Basque Foundation for Science, 48009 Bilbao, Spain}
\author{Daniele De Martino}
\affiliation{Biofisika Institute (CSIC, EHU), 48940 Leioa, Spain}
\affiliation{IKERBASQUE, Basque Foundation for Science, 48009 Bilbao, Spain}


\begin{abstract}
We develop a dynamical mean-field theory for nonreciprocal associative networks that store an extensive number of dynamical attractors, from limit cycles to strange attractors.
Using a path integral calculation under quenched disorder, we derive self-consistent dynamical mean-field equations for pattern overlaps, autocorrelations and response functions.
Memory retrieval capacity is governed by the spectral structure of the coupling matrices encoding stored patterns.
When their eigenvalues are coherently aligned, retarded self-interactions and quenched noise feed back destructively: at zero eigenphase (fixed point attractors) the classical equilibrium capacity bound is recovered, while for limit cycles retrieval collapses far below it.
In contrast, for uniformly distributed eigenphases, retarded self-interactions and much of the quenched noise cancels, reducing the dynamics to an effective single-spin process and amplifying capacity substantially.
We validate the theory against microscopic simulations for limit-cycle and chaotic attractors, identifying eigenvalue decoherence as the mechanism enabling enhanced storage of dynamical memories.
\end{abstract}

\maketitle

Biological neural circuits sustain multiple oscillatory rhythms simultaneously, spanning distinct timescales and coexisting within the same circuits~\cite{colgin2016rhythms,buzsaki2023comeofage}.
Such multiplicity is a long-standing theme in systems neuroscience \cite{buzsaki2006rhythms,wang2010neurophysiological},
 proposed to support a  code in which different frequency bands carry distinct information streams \cite{lisman2013theta}.
Statistical mechanics has proven fundamental to understanding how neural network models function as associative memories~\cite{amit1989modeling,coolen2005theory}, but 
it remains an open question how a recurrent network can simultaneously store an extensive number of distinct dynamical attractors, each with its own temporal structure.
Analytical studies of associative memory have largely focused on the storage of static patterns~\cite{hopfield1982neural,amit1985storing}, with symmetric interactions among neuronal units.
Nonreciprocal (asymmetric) couplings extend this to sequential
association~\cite{amari1972learning,sompolinsky1986temporal,kleinfeld1986sequential,kleinfeld1988associative,gutfreund1988processing} and oscillations \cite{avni2025nonreciprocal,avni2025dynamical},  but the capacity of such networks has been characterized only for a single stored sequence or cycle~\cite{during1998phase,kawamura2002transient} and nonreciprocal diluted systems retrieving static patterns \cite{derrida1987exactly,derrida1987learning,crisanti1988dynamics}.
Similar nonreciprocal Hebbian structures underlie linear attention mechanisms~\cite{schlag2021linear}. Exponential hetero-associative networks have become an active object of study to characterise nonlinear attention~\cite{lucibello2024exponential,agliari2026exponential},
connecting the memory capacity of sequential memories to modern transformer architectures. 

Here we develop a  a dynamical mean-field theory for nonrecirocal associative memories that encode an extensive number of  dynamical attractors with distinct frequencies and structure, and we show that the retrieval capacity is governed by the spectral structure of the couplings between encoded patterns.

\PRLsection{Model}
We consider a network of $N$ binary spins,
$\bm x_t=(x_{1,t},\ldots,x_{N,t})$, with
$x_{i,t}\in\{\pm1\}$ and $t=0,\ldots,\mathcal T$. Spins are updated stochastically in
discrete time, with update probability $\Delta$, driven
by effective fields $\bm h_t=(h_{1,t},\ldots,h_{N,t})$ defined as
\begin{equation}
    h_{i,t}=H_{i,t}+\sum_{j; j\neq i} J_{ij}x_{j,t},
\end{equation}
where $H_{i,t}$ are time-dependent external fields.

Couplings $J_{ij}$, with $J_{ii}=0$, are constructed from $1+L$ blocks of interacting random patterns.
Each block $\upsilon$ encodes one dynamical attractor, each containing $M_\upsilon$ patterns
$\bm\xi_\upsilon^a=(\xi_{1,\upsilon}^a,\ldots,\xi_{N,\upsilon}^a)$. Patterns are composed by  i.i.d. random variables
 $\xi_{i,\upsilon}^a\in\{\pm1\}$, coupled through an
$M_\upsilon\times M_\upsilon$ matrix $\bm A^\upsilon$:
\begin{equation}
    J_{ij}
    =\frac{1}{N}\sum_{\upsilon=0}^{L}
      \sum_{a,b=1}^{M_\upsilon}
      A_{ab}^\upsilon\xi_{i,\upsilon}^a\xi_{j,\upsilon}^b,
    \qquad i\neq j.
    \label{eq:Jij}
\end{equation}

Trajectories $\bm x=\{\bm x_0,\dots,\bm x_{\mathcal T}\}$ evolve through stochastic
updates governed by independent variables $\tau_{i,t}\sim\mathrm{Bernoulli}(\Delta)$,
which determine whether spin $i$ is updated at time $t$ via Glauber dynamics at
inverse temperature $\beta=1/T$ ($\tau_{i,t}=1$), or held fixed ($\tau_{i,t}=0$). For a
given field history $\bm h=\{\bm h_0,\dots,\bm h_{\mathcal T}\}$, this defines the
path probability
\begin{align}
    p_{\bm h}(\bm x, \bm \tau) =&   \prod_{i,t} p(\tau_{i,t+1}) \big( (1-\tau_{i,t+1}) \delta_{x_{i,t+1},x_{i,t}} 
    \nonumber\\ & + \tau_{i,t+1} \frac{1}{2} (1+ x_{i,t} \tanh[\beta h_{i,t}]) \big).
\end{align}
The parameter $\Delta$ interpolates between parallel dynamics ($\Delta=1$) and single–spin updates ($\Delta\to 0$); for all numerical results in this Letter we use an intermediate value $\Delta=0.1$.
Note that leaving $\bm h$ as a generic function allows us to later introduce
the self-consistent fields from our mean-field theory.


\PRLsection{Quenched disorder and order parameters}
We single out block patterns for $\upsilon=0$ as the target sequential memory, and split the patterns into this main target set and the remaining $P:=\sum_{\upsilon=1}^{L}M_\upsilon$ sets as a source of quenched disorder,
\begin{align}
    J_{ij}
    &= \frac{1}{N} \bm\xi_i^\top \bm A \bm\xi_j
       + \frac{1}{N} \bm{\hat{\xi}}_i^\top \bm{\hat{A}} \bm{\hat{\xi}}_j, \qquad i \neq j, 
    \label{eq:Jij2}
\end{align}
with the target pattern $\bm \xi_i \coloneqq \bm \xi_{i,0}$, $\bm A \coloneqq \bm A^0$ (with $M=M_0$), and joint patterns $\bm{\hat{\xi}}_i \coloneqq (\bm \xi_{i,1},\dots, \bm \xi_{i,P})$ and  couplings $\bm{\hat{A}} \coloneqq \mathrm{diag}(\bm A^1, \ldots, \bm A^L)$. The memory load per spin is defined as $\alpha = P/N$.

\PRLsection{Dynamical mean-field theory}
We introduce a moment-generating functional for trajectories, for each realization of $\bm{\hat{\xi}}$, adding fields $\bm g$:
\begin{align}
    Z_{\bm{\hat{\xi}}}(\bm g)
    &= \ang{
    e^{\sum_{i,t} x_{i,t} g_{i,t}} }_{\bm h},
    \label{eq:ZT}
\end{align}
with  $\langle f(\bm x)\rangle_{\bm h} = \sum_{\bm x, \bm \tau}  p_{\bm h}(\bm x, \bm \tau) f(\bm x)$.
Derivatives of $Z_{\bm{\hat{\xi}}}(\bm g)$ retrieve spin moments, e.g., $\partial Z_{\bm{\hat{\xi}}}(\bm 0)/\partial g_{i,t} = \ang{x_{i,t}}_{\bm h}$, and response functions, $ {\partial^2 Z_{\bm{\hat{\xi}}}(\bm 0)}/(\partial g_{i,t}\partial H_{j,s}) = {\partial \ang{x_{i,t}}_{\bm h}}/{\partial H_{j,s}}$.

We define dynamical order parameters for the pattern overlaps,
correlations, and response functions:
\begin{align}
    m_t^a &= \frac{1}{N}\sum_i \xi_i^a  \frac{\partial Z_{\bm{\hat{\xi}}}(\bm 0)}{\partial g_{i,t}}  =  \frac{1}{N}\sum_i \xi_i^a \ang{x_{i,t}}_{\bm h},
    \label{eq:m_empirical}\\
    q_{t,s} &= \frac{1}{N}\sum_i \frac{\partial^2 Z_{\bm{\hat{\xi}}}(\bm 0)}{\partial g_{i,t}\partial g_{i,s}} =  \frac{1}{N}\sum_i \ang{x_{i,t}x_{i,s}}_{\bm h},
    \label{eq:q_empirical}\\
    \chi_{t,s} &= \frac{1}{N}\sum_i \frac{\partial^2 Z_{\bm{\hat{\xi}}}(\bm 0)}{\partial g_{i,t}\partial H_{i,s}}
    \nonumber\\ &= \frac{\beta}{N}\sum_i \ang{x_{i,t} \tau_{i,s+1}(x_{i,s+1} - \tanh[\beta h_{i,s}])}_{\bm h},
    \label{eq:varrho_empirical}
\end{align}
We also introduce $\rho_{t,s}= N^{-1} \sum_i{\partial^2 Z_{\bm{\hat{\xi}}}(\bm 0)}/(\partial H_{i,t}\partial H_{i,s})$, which becomes equal to zero as $\partial Z_{\bm{\hat{\xi}}}(\bm 0)/\partial H_{i,t}=0$.

To obtain closed equations for the order parameters, we follow a path-integral approach~\cite{sommers1987path,coolen2005theory} to calculate the average $\langle\langle{Z_{\bm{\hat{\xi}}}(\bm g)}\rangle\rangle = \sum_{\bm{\hat{\xi}}} p(\bm{\hat{\xi}}) Z_{\bm{\hat{\xi}}}(\bm g)$ over  uniformly distributed independent patterns $\xi_{i,\upsilon}^a$ for $\upsilon>0$ so $ p(\bm{\hat{\xi}}) = 2^{-P}$. We introduce the order parameters above plus additional auxiliary variables via integral representations of delta functions to allow averaging over quenched disorder  (see SM \cite{supplementalmaterial} for a detailed calculation, a summary of which is reported in  \emph{End matter}). Finally saddle-point evaluation results in the quenched generating functional
\begin{align}
    \label{eq:quenched_generating_functional}
	\langle\langle{Z_{\bm{\hat{\xi}}}(\bm g)}\rangle\rangle
    \propto & \int d\bm z p(\bm z)
    \frac{1}{2^M}\sum_{\bm\sigma}
       \ang{e^{\sum_{i,t}x_{i,t} g_{i,t}}}_{\bm\hmf^{\bm\sigma}}.
 \\ \hmf_{i,t}^{\bm\sigma} &= H_{i,t} + \sum_{ab}\sigma_a A_{ab} m_t^b + \alpha\sum_{s; s<t} K_{t,s} x_{i,s} + z_{i,t}.
\end{align}
Here,  we represent the realizations of $\bm{\xi}_i = (\xi_i^1,\ldots,\xi_i^M)$ by a binary vector $\bm{\sigma}\in\{\pm1\}^M$ with $2^M$ possible states (with $M=M_0$).
The term  $\bm z_i = (z_{i,0}, \dots, z_{i,\mathcal T})$ is a Gaussian variable determined by $p(\bm z_i) \sim \mathcal{N}(\bm 0, \alpha\bm R)$.
The result relies on the kernels $\bm R, \bm K$, defined as functions of $\bm q, \bm \chi$ and $\bm{\hat{A}}$
\begin{align}
    R_{t,s}  &= \frac{1}{P}
       \sum_{a=0}^{P-1}
       \big[\bm\kappa^{-1}
           \big(\bm I_P \otimes \bm q\big)
           \bm\kappa^{-\top}
       \Big]_{(aT+t,aT+s)},
       \label{eq:R_saddle}
    \\
    K_{t,s}
    &= \frac{1}{P}
       \sum_{a=0}^{P-1}
       \big[-\bm\kappa^{-1}
       \big]_{(aT+t,aT+s)},
    \\ \bm\kappa & \coloneqq  \bm I_P\otimes\bm\chi - \bm{\hat{A}}^{-1}\otimes\bm I_{\mathcal T}.
    \label{eq:K_saddle}
\end{align}
being $\otimes$ the Kroneker product and 
where $\bm\kappa^{-\top} = (\bm\kappa^{\top})^{-1}$. 
Note that the diagonal exclusion $J_{ii}=0$ generates an Onsager correction that cancels $K_{t,t}$ exactly (see \emph{End Matter}), so the retarded kernel couples $x_{t+1}$ only to spins at $s<t$.
The result above can be simplified in the case of an orthonormal matrix $\bm{\hat{A}}$ with eigenvalues $\{\hat\lambda_r\}$ yielding
\begin{align}
    \bm R  &= \frac{1}{P}
       \sum_{n,m=0}^\infty
       \sum_{r=1}^{P}
       \hat\lambda_r^{n+1}(\hat\lambda_r^\ast)^{m+1}
           \bm\chi^n \bm q (\bm\chi^\top)^m,
\label{eq:R_eigenvalues}
\\  \bm K  &=\frac{1}{P} \sum_{n=0}^\infty \sum_{r=1}^P \hat\lambda_r^{n+1}      \bm\chi^n.
\label{eq:K_eigenvalues}
\end{align}
As $\bm\chi$ is strictly lower diagonal, Eqs.~(\ref{eq:R_eigenvalues}-\ref{eq:K_eigenvalues}) can be computed at time $t+1$ using only causal influences from time $s\le t$ (see Eqs.~\ref{eq:recursion_G_general}-\ref{eq:recursion_S_general} in the \emph{End Matter}).
Furthermore, the equation above results in diverging $\bm R, \bm K$ when the spectral radius of $\bm\chi$ is larger than 1. This predicts that memory retrieval will be unfeasible beyond that point (quenched noise is amplified indefinitely).

\PRLsection{Mean-field equations}
The quenched generating functional~\eqref{eq:quenched_generating_functional}
factorises over spins once the kernels $\bm R$ and $\bm K$ are fixed.
 Dropping $i$ terms for simplicity, for zero fields and orthonormal $\bm{\hat{A}}$, we describe the behaviour of the system with simplified mean-field equations
\begin{align}
\bm m_{t} &= \frac{1}{2^M}\sum_{\bm\sigma} \bm\sigma \ang{x_{t}}_{\bm\hmf^{\bm\sigma}}
\label{eq:mean-field-dynamics-patterns}
\\ \chi_{t,s} &= \frac{\beta}{2^M}\sum_{\bm\sigma} 
 \ang{x_{t} \tau_{s+1}(x_{s+1} - \tanh[\beta\hmf_s])}_{\bm\hmf^{\bm\sigma}},
\label{eq:mean-field-dynamic-varrho}
\end{align}


Two-time correlations propagate recursively. For $t>s$,
\begin{align}
    q_{t+1,s}
    &=(1-\Delta)q_{t,s}
      +\ang{\tau_{t+1}x_{t+1}x_s}_
       {\bm{\bar h}^{\bm\sigma}},
    \label{eq:q_step1}
    \\
    \ang{\tau_{t+1}x_{t+1}x_{s+1}}_
    {\bm{\bar h}^{\bm\sigma}}
    &=(1-\Delta)
      \ang{\tau_{t+1}x_{t+1}x_s}_
      {\bm{\bar h}^{\bm\sigma}}
      \nonumber\\
    &\quad+
      \ang{\tau_{t+1}\tau_{s+1}
      x_{t+1}x_{s+1}}_
      {\bm{\bar h}^{\bm\sigma}} .
\end{align}
and, for $t\neq s$, 
\begin{equation}
    \ang{\tau_{t+1}\tau_{s+1}x_{t+1}x_{s+1}}_
    {\bm{\bar h}^{\bm\sigma}}
    =
    \frac{\Delta^2}{2^M}
    \sum_{\bm\sigma}
    \ang{
      \tanh(\beta\bar h_t^{\bm\sigma})
      \tanh(\beta\bar h_s^{\bm\sigma})
    }_{\bm{\bar h}^{\bm\sigma}}.
    \label{eq:sigma_sigma_mf}
\end{equation}

Solving these equations is made intractable by the self-dependence of $\bm x$ through couplings $\bm K$ and the non-Markovian history encoded in $\bm R$. The mean field behaviour can be estimated by Monte Carlo sampling of the effective single-spin process (similarly to \cite{eissfeller1992new}), or directly solving the equations in special cases where $\bm K=\bm 0$.

\begin{figure*}
    \centering
    \begin{tikzpicture}
        \node[inner sep=0, anchor=north west] (mainfigA) at (0,0) {%
            \includegraphics[width=0.33\textwidth]{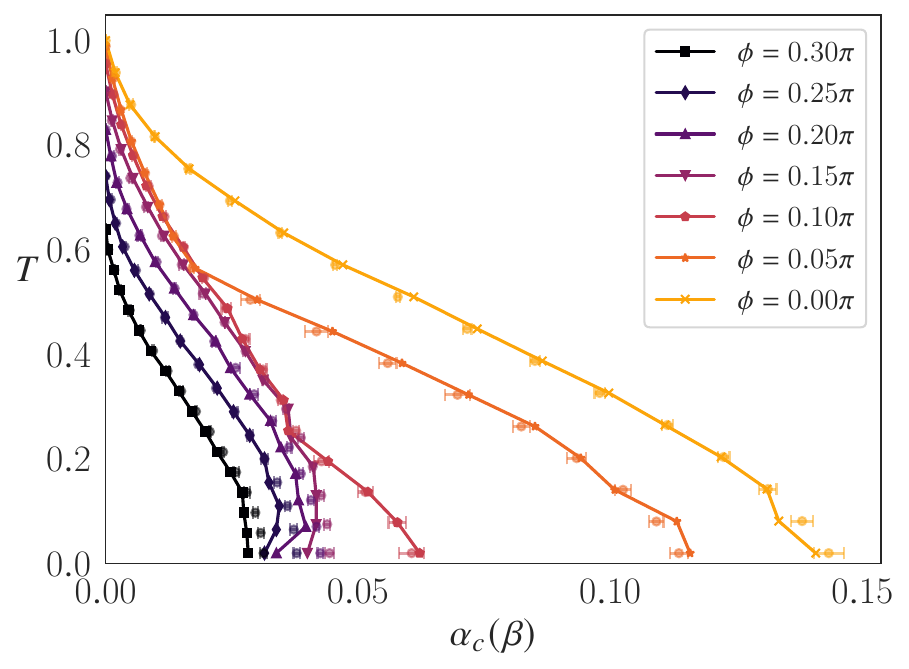}
        };
        \node[
            anchor=north west,
            inner sep=0
        ] at ([xshift=2mm]mainfigA.north east) (eigA) {%
            \includegraphics[width=0.12\textwidth]{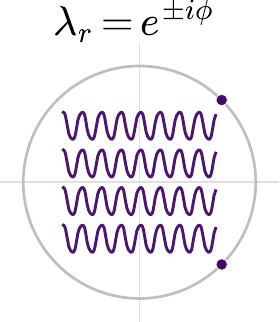}
        };
        \node[anchor=north west, xshift=-3mm] at (mainfigA.north west) {(a)};

        \node[inner sep=0, anchor=north west] (mainfigB) at ([xshift=8mm]eigA.north east) {%
            \includegraphics[width=0.33\textwidth]{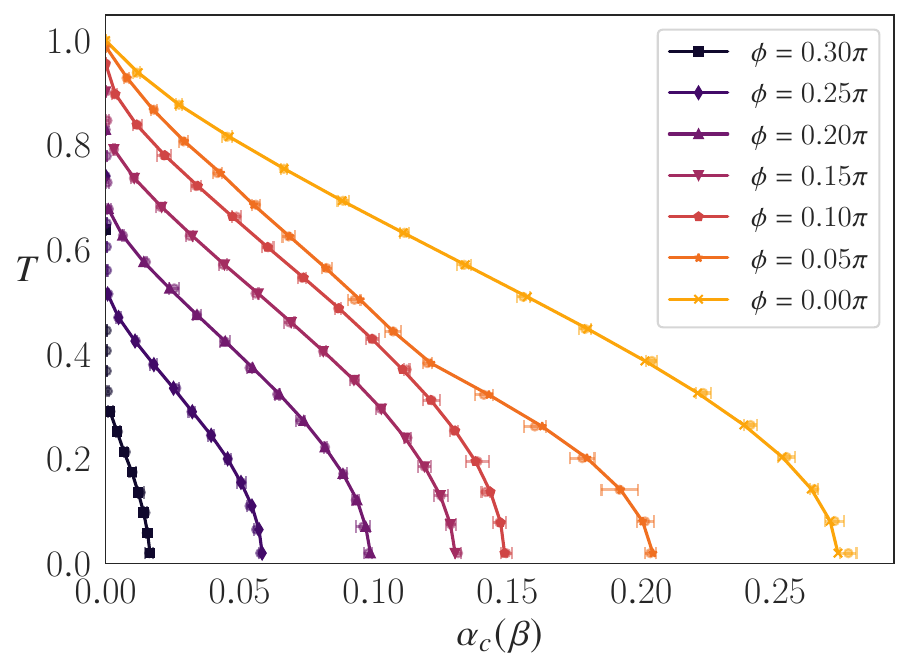}
        };
        \node[
            anchor=north west,
            inner sep=0
        ] at ([xshift=2mm]mainfigB.north east) {%
            \includegraphics[width=0.12\textwidth]{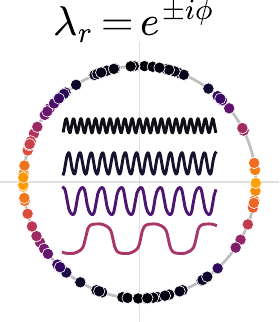}
        };
        \node[anchor=north west, xshift=-3mm] at (mainfigB.north west) {(b)};
    \end{tikzpicture}
    \vspace{-2mm}
    \caption{Critical capacity $\alpha_c$ as a function of temperature $T=1/\beta$ for rotation matrices with eigenvalues $\lambda_r=e^{\pm\iu\phi}$.
    \textbf{a)} All submatrices have coherently aligned eigenphases $\phi_r=\phi$. 
    \textbf{b)} Uniformly distributed eigenphases $\phi_r\sim\mathcal{U}[0,2\pi]$, with $\phi$  the eigenphase of the target $\bm A^0$ block.
    Error bars denote the standard deviation over trials of the microscopic network at $N=500,000$.}
    \label{fig:Fig1}
\end{figure*}

\PRLsection{Encoding identical cycles}
Since matrices $\bm A^\upsilon$ are real, their eigenvalues come in conjugate pairs $e^{\pm\iu\phi_r}$ (with an additional $\pm1$ eigenvalue when the dimension is odd).
We first consider the simplest case of identical $2{\times}2$ rotation matrices $\bm A^\upsilon=\bm A = \bm \Omega_\phi$, where $\bm \Omega_\phi \coloneqq \begin{bsmallmatrix} \cos\phi & \sin\phi\\ -\sin\phi & \cos\phi\end{bsmallmatrix}$. 
This yields
\begin{align}
    \bm R  &= \sum_{n,m=0}^\infty \cos((n-m)\phi) \bm\chi^n \bm q (\bm\chi^\top)^m,
\\  \bm K  &= \sum_{n=0}^\infty \cos((n+1)\phi)\bm\chi^n.
\end{align}
Decomposing $\bm R = \operatorname{Re}[\bm S]$ and $\bm K = \operatorname{Re}[\bm G]$ with complex auxiliary matrices, the order parameters satisfy $\bm S = \bm q + e^{\iu\phi}\bm\chi\bm S + e^{-\iu\phi}\bm S\bm\chi^\top - \bm\chi\bm S\bm\chi^\top,  \bm G = e^{\iu\phi}\!\left(\bm I + \bm\chi\bm G\right) $
which can be solved forward in time exploiting the strictly lower-triangular (causal) structure of $\bm\chi$.

The rotation angle $\phi$ interpolates between qualitatively different dynamical regimes.
For $\phi=0$ the attractors are fixed points, recovering the results of a classical Hopfield network (similar to \cite{amit1985storing}) but for partially parallel dynamics.
As $\phi$ increases from zero the network can sustain genuine cyclic orbits, and the critical capacity drops sharply.
Already at zero load the dynamics undergoes a heteroclinic bifurcation at $\beta^*(\phi)$, where the cycle slows near four saddle-type fixed points until its period diverges and the orbit collapses onto fixed points (see \emph{End Matter}; see also Ref.~\cite{xue2025critical}).
This boundary also appears in Figure~\ref{fig:Fig1}(a): warm-colored curves lie on the fixed-point side of $\beta^*$ and decay slowly, cooler ones retrieve limit cycles.
Once oscillations emerge, the retarded self-interaction $\bm K$ and the cross-terms in $\bm R$ feed back destructively, drastically reducing the number of storable patterns.

\PRLsection{Encoding cycles with isotropic eigenvalues}
A remarkable simplification occurs when the eigenvalues of $\bm{\hat{A}}$ are uniformly distributed on the unit circle.
The sums over eigenvalues in \eqref{eq:K_eigenvalues} and over conjugate pairs in \eqref{eq:R_eigenvalues} are equal to zero except the latter when $n=m$, resulting in
\begin{equation}
    \bm R  = \sum_{n=0}^\infty \bm\chi^n \bm q (\bm\chi^\top)^n, 
    \qquad \bm K = 0.
\end{equation}
The new form allows us to write $\bm R$ in the form of the Lyapunov equation
$\bm R=\bm q+\bm\chi \bm R \bm\chi^\top$, and the absence of $\bm K$ eliminates the delayed self-coupling rendering the effective single-spin dynamics equivalent to a nonequilibrium spin glass driven by coloured Gaussian noise. 


For uniform eigenvalues on the unit circle, the stochastic update of the mean-field equations reads
\begin{equation}
\bm m_{t+1}=(1-\Delta)\bm m_t +\frac{\Delta}{2^M}\sum_{\bm\sigma} \bm\sigma\int Dz_t
\tanh[\beta\hmf_t^{\bm\sigma}],
\label{eq:mf_m_async_app}
\end{equation}
with susceptibility, for $t>s$,
\begin{equation}
\chi_{t,s}=\Delta (1-\Delta)^{t-s-1} \frac{\beta}{2^M}\sum_{\bm\sigma}\int Dz_s
\Big(1-\tanh^2[\beta\hmf_s^{\bm\sigma}]\Big),
\label{eq:kappa_def_async_app}
\end{equation}
where $Dz=dz\frac{1}{\sqrt{2\pi}}e^{-\frac{1}{2}z^2}$.

In this case, we can solve the self-consistent equations directly without recurring to Monte Carlo methods. 
Figure~\ref{fig:Fig1}(b) shows the resulting phase boundary $\alpha_c$ for 
independent rotation matrices $\bm A^\upsilon = \bm \Omega_{\phi_\upsilon}$ with angles $\phi_\upsilon$ sampled uniformly from $[0,2\pi)$, so that each stored attractor oscillates at a different, randomly chosen frequency.
The $\phi=0$ curve shows $\alpha_c(T=0)\simeq 0.27$, roughly twice the classical Hopfield capacity of $0.138$. This parallels the result in \cite{during1998phase} for a single infinite-length sequence at $\Delta = 1$, whose solution is equivalent to the equilibrium solution of the system at $\phi=0$ for any $\Delta$. All other curves interpolate smoothly toward it, suggesting the destructive feedback responsible for the capacity collapse in Figure~\ref{fig:Fig1}(a) is an artifact of coherent eigenphase alignment rather than an intrinsic cost of storing cyclic attractors.

Compared with the single-angle case (Figure~\ref{fig:Fig1}(a)), the memory capacity is greatly amplified: for all angles the capacity substantially exceeds that of the corresponding limit-cycle regime. The exception is the edge case $\phi=0.3\pi$. We find attractors beyond $\phi\approx   0.32\pi$ cannot be stored at all, and memory retrieval fails even for infinitely small $\alpha$. This is predicted nicely by our theory. For $\alpha=0$, we find for large $\beta$ a transition (see Figure \ref{fig:Fig4}) where the spectral value of $\bm\chi$ becomes larger than 1, making $\bm R$ diverge for large t. Thus, near this limit the memory capacity is reduced instead of amplified.  

\begin{figure}
    \noindent
    \begin{tikzpicture}
        \node[anchor=north west, inner sep=0pt] (b)
            at (0, 0)
            {\includegraphics[width=0.8\linewidth]{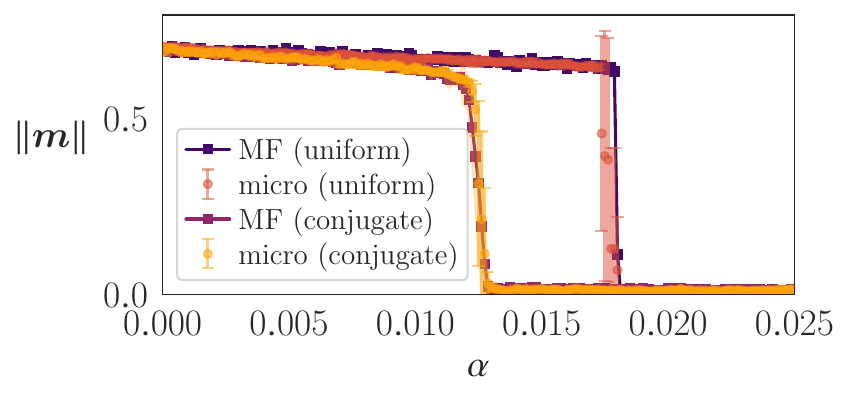}};
        \node at ([xshift=8pt, yshift=-4pt]b.north west) {(b)};
        \node[anchor=north east, inner sep=4pt, fill=white, draw=black]
            (a)
            at ([xshift=45pt, yshift=32pt]b.north east)
            {\includegraphics[width=0.32\linewidth,
                trim=80pt 15pt 50pt 20pt, clip]{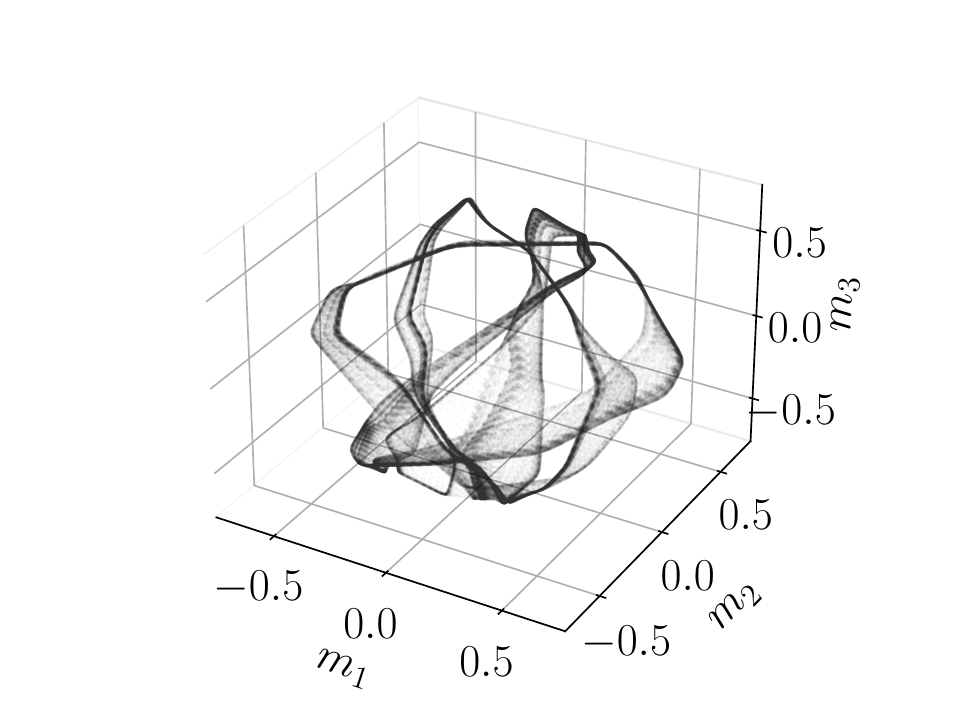}};
        \node at ([xshift=8pt, yshift=-8pt]a.north west) {(a)};
    \end{tikzpicture}%
    \vspace{-2mm}
    \caption{Chaotic attractor retrieval.
    (a)~Strange attractor of the $\alpha=0$ overlap dynamics for a
    $4{\times}4$ orthogonal matrix $\bm A$, projected onto
    $(m_1,m_2,m_3)$.
    (b)~Overlap norm $\|\bm m\|$ versus memory load $\alpha$ at
    $\beta=14.73$, $\Delta=0.1$, comparing the mean-field theory
    (MF) with microscopic simulations (micro) for
    conjugate and uniform eigenvalue distributions.
    Shaded bands and error bars denote standard deviation over trials.
    }
    \label{fig:Fig2}
\end{figure}

\PRLsection{Retrieval of a chaotic attractor}
The framework extends naturally to higher-dimensional coupling matrices.
For $M = 4$, a real orthogonal matrix $\bm A$ has two conjugate eigenvalue pairs $e^{\pm\iu\phi_1}$, $e^{\pm\iu\phi_2}$. 
For suitable choices of $\bm A$ this orbit can become quasiperiodic or chaotic. We used a genetic algorithm (see SM) to find matrices $\bm A$ resulting in chaotic dynamics for $\alpha=0$ and spectral radius of $\bm\chi$ smaller than one to ensure the attractor can be stored in memory. 
Figure~\ref{fig:Fig2}(a) shows a strange attractor obtained by this procedure, with $\phi_1\approx 0.23\pi$, $\phi_2\approx 1.73\pi$, and Lyapunov exponents $\lambda_1\approx 0.032$, $\lambda_2\approx -0.001$, projected onto the first three pattern overlaps.

Figure~\ref{fig:Fig2}(b) shows the memory capacity of this chaotic attractor for the conjugate and uniform eigenvalue distributions, respectively (see \emph{End matter}).
In both cases the mean-field prediction agrees quantitatively with microscopic simulations, validating the mean-field description of chaotic attractor retrieval under quenched disorder.
As in the limit-cycle regime, the uniform distribution yields a higher critical load than the conjugate one, supporting the hypothesis that eigenvalue decoherence amplifies memory capacity independently of the attractor topology.


\begin{figure}
    \centering
    \includegraphics[width=0.8\linewidth]{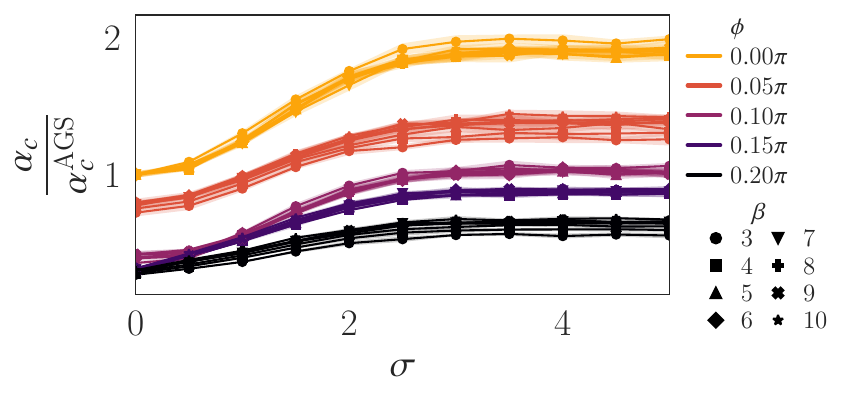}

    \vspace{-6mm}
    \caption{
    Normalised memory capacity as a function of the eigenphase standard deviation $\sigma$
    around a central angle $\phi$, for several values of $\phi$.
    Shaded bands denote standard deviation over trials.
    }
    \label{fig:Fig3}
\end{figure}

\PRLsection{Partial coherence}
The results above suggest that it is the \emph{diversity} of eigenphases across stored attractors that controls the memory capacity. To test this directly, we come back to the case of storing 2D limit cycles, this time by drawing each eigenphase from a normal distribution centred at a common angle $\phi$ with standard deviation $\sigma$, so that $\sigma=0$ recovers the fully coherent case of Figure~\ref{fig:Fig1}(a) and increasing $\sigma$ progressively decorrelates the phases across attractors,. 
through numerical simulations with $N=50,000$ spins. Figure~\ref{fig:Fig3} shows the resulting critical capacity $\alpha_c(\beta,\sigma,\phi)$ for several central angles $\phi$ and inverse temperatures $3\leq\beta\leq 10$, normalized by the classical memory capacity solution $\alpha^\mathrm{AGS}(\beta) = \alpha_c(\beta,0,0)$ at $\phi=0$ \cite{amit1985storing} and a factor $f(\phi)$ minimising the mean squared error between $\alpha_c(\beta,0,\phi)$ and $\alpha^\mathrm{AGS}(\beta)f(\phi)$. For all but the largest $\phi$, capacity increases monotonically with $\sigma$, improving memory capacity by a factor between 1.5 and 3.  Note that large values of $\phi$, too close to the instability boundary, do not display a memory enhancement with phase decoherence.

\PRLsection{Discussion}
Recurrent networks with nonreciprocal Hebbian couplings can store an extensive number of dynamical attractors, provided the response matrix $\bm\chi$ has a spectral radius smaller than one. The resulting memory capacity crucially depends on the spectral structure of the encoding matrices $\bm{\hat{A}}$.
Since coherent eigenphases make the cross-terms in $\bm R$ and the retarded couplings $\bm K$ feed back destructively while decoherence cancels both, it is the spectral diversity of the stored attractors, rather than their dynamical topology, that sets how many of them can coexist.
Both quantities also simplify by other mechanisms. In diluted networks~\cite{derrida1987exactly, coolen2005theory}, $\bm R$ collapses onto the spin autocorrelation, while $\bm K$ is also eliminated for fully asymmetric dilutions. As in the case of eigenvalue decoherence, dilution severs crosstalk interactions among patterns.

From a biological perspective, the coexistence of multiple oscillatory modes with incommensurate frequencies is ubiquitous in hippocampal and cortical circuits~\cite{penttonen2003logarithmic,buzsaki2023comeofage}, with different bands carrying different information streams~\cite{colgin2009frequency}.
Our results give this spectral diversity a functional rationale: attractors spanning incommensurate frequencies interfere less, so capacity grows with the spread of eigenphases.
In the more general context of complex systems dynamics the fact that diversity amplifies rather than degrades capacity runs against the intuition, formalised by May~\cite{may1972will,may1973stability}, that complexity destabilises large interacting systems. 
Analogous inversions appear elsewhere: from heterogeneous excitability tuning stability in  excitatory-inhibitory neural networks~\cite{hutt2023neural} to diversity of species in generalized Lotka-Volterra models with sublinear population growth promoting stability~\cite{hatton2024diversity}.
In each case, a form of diversity (eigenphases, excitability thresholds, or number of species) turns increasing complexity from a source of instability into a stabilising force.

\section*{Acknowledgements}
\begin{acknowledgments}
M.A. is partly supported by the Ramón y Cajal programme (RYC2022-038204-I), funded by MICIU/AEI/10.13039/501100011033 and by the European Social Fund Plus (FSE+), Grant PID2023-146869NA-I00 funded by MICIU/AEI/10.13039/501100011033 and cofunded by the European Union, and Basque Government ELKARTEK funding (code KK-2023/00085). He is also supported by the Basque Government through the BERC 2022-2025 program and by the Spanish State Research Agency through BCAM Severo Ochoa excellence accreditation CEX2021-01142-S funded by MICIU/AEI/10.13039/501100011033.
 D.D.M. acknowledges financial support from the grants PIBA 2024 1 0016 (Basque Government) and Project PID2023-146408NB-I00 funded by MICIU/AEI/10.13039/501100011033 and by FEDER, UE.
\end{acknowledgments}

\section*{Author Contributions}
M.A.\ and D.D.M.\ conceived the study.
M.A.\ developed the analytical theory and wrote the code implementing it.
Both authors contributed to the numerical simulations, interpreted the results, and wrote the manuscript.

\section*{Data Availability Statement}

The code used in this study is available in a public repository~\cite{coderepo}.

\bibliography{references}

@misc{agliari2026exponential,
  author        = {Agliari, Elena and Barra, Adriano and Ladiana, Andrea and Lepre, Andrea},
  title         = {Exponential Capacity in Multilayer Hetero-Associative Neural Networks},
  year          = {2026},
  eprint        = {2607.29554},
  archivePrefix = {arXiv},
  doi           = {10.48550/arXiv.2607.29554}
}

@article{amari1972learning,
  author  = {Amari, Shun-Ichi},
  title   = {Learning patterns and pattern sequences by self-organizing nets of threshold elements},
  journal = {IEEE Trans. Comput.},
  volume  = {C-21},
  number  = {11},
  pages   = {1197--1206},
  year    = {1972},
  doi     = {10.1109/T-C.1972.223477}
}

@article{amit1985storing,
  author  = {Amit, Daniel J. and Gutfreund, Hanoch and Sompolinsky, Haim},
  title   = {Storing infinite numbers of patterns in a spin-glass model of neural networks},
  journal = {Phys. Rev. Lett.},
  volume  = {55},
  number  = {14},
  pages   = {1530--1533},
  year    = {1985},
  doi     = {10.1103/PhysRevLett.55.1530}
}

@book{amit1989modeling,
  author    = {Amit, Daniel J.},
  title     = {Modeling Brain Function: The World of Attractor Neural Networks},
  publisher = {Cambridge University Press},
  address   = {Cambridge},
  year      = {1989}
}

@article{avni2025dynamical,
  author  = {Avni, Yael and Fruchart, Michel and Martin, David and Seara, Daniel and Vitelli, Vincenzo},
  title   = {Dynamical phase transitions in the nonreciprocal {I}sing model},
  journal = {Phys. Rev. E},
  volume  = {111},
  pages   = {034124},
  year    = {2025},
  doi     = {10.1103/PhysRevE.111.034124}
}

@article{avni2025nonreciprocal,
  author  = {Avni, Yael and Fruchart, Michel and Martin, David and Seara, Daniel and Vitelli, Vincenzo},
  title   = {Nonreciprocal {I}sing Model},
  journal = {Phys. Rev. Lett.},
  volume  = {134},
  pages   = {117103},
  year    = {2025},
  doi     = {10.1103/PhysRevLett.134.117103}
}

@book{buzsaki2006rhythms,
  author    = {Buzs\'{a}ki, Gy\"{o}rgy},
  title     = {Rhythms of the Brain},
  publisher = {Oxford University Press},
  address   = {New York},
  year      = {2006}
}

@article{buzsaki2023comeofage,
  author  = {Buzs\'{a}ki, Gy\"{o}rgy and V\"{o}r\"{o}slakos, Mih\'{a}ly},
  title   = {Brain rhythms have come of age},
  journal = {Neuron},
  volume  = {111},
  number  = {7},
  pages   = {922--926},
  year    = {2023},
  doi     = {10.1016/j.neuron.2023.03.018}
}

@misc{coderepo,
  author = {Aguilera, Miguel},
  title  = {{Storing Infinite Dynamical Attractors in Nonreciprocal Associative Neural Networks}: Code repository},
  year   = {2026},
  url    = {https://github.com/MiguelAguilera/DynamicalMemoryCapacity}
}

@article{colgin2009frequency,
  author  = {Colgin, Laura Lee and Denninger, Tobias and Fyhn, Marianne and Hafting, Torkel
             and Bonnevie, Tora and Jensen, Ole and Moser, May-Britt and Moser, Edvard I.},
  title   = {Frequency of gamma oscillations routes flow of information in the hippocampus},
  journal = {Nature},
  volume  = {462},
  pages   = {353--357},
  year    = {2009},
  doi     = {10.1038/nature08573}
}

@article{colgin2016rhythms,
  author  = {Colgin, Laura Lee},
  title   = {Rhythms of the hippocampal network},
  journal = {Nat. Rev. Neurosci.},
  volume  = {17},
  number  = {4},
  pages   = {239--249},
  year    = {2016},
  doi     = {10.1038/nrn.2016.21}
}

@book{coolen2005theory,
  author    = {Coolen, A. C. C. and K\"{u}hn, Reimer and Sollich, Peter},
  title     = {Theory of Neural Information Processing Systems},
  publisher = {Oxford University Press},
  address   = {Oxford},
  year      = {2005}
}

@article{crisanti1988dynamics,
  author  = {Crisanti, A. and Sompolinsky, H.},
  title   = {Dynamics of spin systems with randomly asymmetric bonds: {I}sing spins and {G}lauber dynamics},
  journal = {Phys. Rev. A},
  volume  = {37},
  pages   = {4865},
  year    = {1988},
  doi     = {10.1103/PhysRevA.37.4865}
}

@article{derrida1987exactly,
  author  = {Derrida, B. and Gardner, E. and Zippelius, A.},
  title   = {An exactly solvable asymmetric neural network model},
  journal = {Europhys. Lett.},
  volume  = {4},
  number  = {2},
  pages   = {167--173},
  year    = {1987},
  doi     = {10.1209/0295-5075/4/2/007}
}

@article{derrida1987learning,
  author  = {Derrida, B. and Nadal, J.-P.},
  title   = {Learning and forgetting on asymmetric, diluted neural networks},
  journal = {J. Stat. Phys.},
  volume  = {49},
  pages   = {993--1009},
  year    = {1987},
  doi     = {10.1007/BF01017556}
}

@article{during1998phase,
  author  = {D\"{u}ring, A. and Coolen, A. C. C. and Sherrington, D.},
  title   = {Phase diagram and storage capacity of sequence processing neural networks},
  journal = {J. Phys. A: Math. Gen.},
  volume  = {31},
  number  = {43},
  pages   = {8607--8621},
  year    = {1998},
  doi     = {10.1088/0305-4470/31/43/005}
}

@article{eissfeller1992new,
  author  = {Eissfeller, H. and Opper, M.},
  title   = {New method for studying the dynamics of disordered spin systems without finite-size effects},
  journal = {Phys. Rev. Lett.},
  volume  = {68},
  number  = {13},
  pages   = {2094--2097},
  year    = {1992},
  doi     = {10.1103/PhysRevLett.68.2094}
}

@article{gutfreund1988processing,
  author  = {Gutfreund, H. and M\'{e}zard, M.},
  title   = {Processing of temporal sequences in neural networks},
  journal = {Phys. Rev. Lett.},
  volume  = {61},
  number  = {2},
  pages   = {235--238},
  year    = {1988},
  doi     = {10.1103/PhysRevLett.61.235}
}

@article{hatton2024diversity,
  author  = {Hatton, Ian A. and Mazzarisi, Onofrio and Altieri, Ada and Smerlak, Matteo},
  title   = {Diversity begets stability: {S}ublinear growth and competitive coexistence across ecosystems},
  journal = {Science},
  volume  = {383},
  number  = {6688},
  pages   = {eadg8488},
  year    = {2024},
  doi     = {10.1126/science.adg8488}
}

@article{hopfield1982neural,
  author  = {Hopfield, John J.},
  title   = {Neural networks and physical systems with emergent collective computational abilities},
  journal = {Proc. Natl. Acad. Sci. U.S.A.},
  volume  = {79},
  number  = {8},
  pages   = {2554--2558},
  year    = {1982},
  doi     = {10.1073/pnas.79.8.2554}
}

@article{hutt2023neural,
  author  = {Hutt, Axel and Rich, Scott and Valiante, Taufik A. and Lefebvre, J\'{e}r\'{e}mie},
  title   = {Intrinsic neural diversity quenches the dynamic volatility of neural networks},
  journal = {Proc. Natl. Acad. Sci. U.S.A.},
  volume  = {120},
  number  = {28},
  pages   = {e2218841120},
  year    = {2023},
  doi     = {10.1073/pnas.2218841120}
}

@article{kawamura2002transient,
  author  = {Kawamura, Masaki and Okada, Masato},
  title   = {Transient dynamics for sequence processing neural networks},
  journal = {J. Phys. A: Math. Gen.},
  volume  = {35},
  number  = {2},
  pages   = {253--266},
  year    = {2002},
  doi     = {10.1088/0305-4470/35/2/306}
}

@article{kleinfeld1986sequential,
  author  = {Kleinfeld, David},
  title   = {Sequential state generation by model neural networks},
  journal = {Proc. Natl. Acad. Sci. U.S.A.},
  volume  = {83},
  number  = {24},
  pages   = {9469--9473},
  year    = {1986},
  doi     = {10.1073/pnas.83.24.9469}
}

@article{kleinfeld1988associative,
  author  = {Kleinfeld, David and Sompolinsky, Haim},
  title   = {Associative neural network model for the generation of temporal patterns:
             {T}heory and application to central pattern generators},
  journal = {Biophys. J.},
  volume  = {54},
  number  = {6},
  pages   = {1039--1051},
  year    = {1988},
  doi     = {10.1016/S0006-3495(88)83041-8}
}

@article{lisman2013theta,
  author  = {Lisman, John E. and Jensen, Ole},
  title   = {The Theta--Gamma Neural Code},
  journal = {Neuron},
  volume  = {77},
  number  = {6},
  pages   = {1002--1016},
  year    = {2013},
  doi     = {10.1016/j.neuron.2013.03.007}
}

@article{lucibello2024exponential,
  author  = {Lucibello, Carlo and M\'{e}zard, Marc},
  title   = {Exponential Capacity of Dense Associative Memories},
  journal = {Phys. Rev. Lett.},
  volume  = {132},
  number  = {7},
  pages   = {077301},
  year    = {2024},
  doi     = {10.1103/PhysRevLett.132.077301}
}

@article{may1972will,
  author  = {May, Robert M.},
  title   = {Will a Large Complex System be Stable?},
  journal = {Nature},
  volume  = {238},
  pages   = {413--414},
  year    = {1972},
  doi     = {10.1038/238413a0}
}

@book{may1973stability,
  author    = {May, Robert M.},
  title     = {Stability and Complexity in Model Ecosystems},
  publisher = {Princeton University Press},
  address   = {Princeton, NJ},
  year      = {1973}
}

@article{penttonen2003logarithmic,
  author  = {Penttonen, Markku and Buzs\'{a}ki, Gy\"{o}rgy},
  title   = {Natural logarithmic relationship between brain oscillators},
  journal = {Thalamus Relat. Syst.},
  volume  = {2},
  number  = {2},
  pages   = {145--152},
  year    = {2003},
  doi     = {10.1016/S1472-9288(03)00007-4}
}

@inproceedings{schlag2021linear,
  author    = {Schlag, Imanol and Irie, Kazuki and Schmidhuber, J\"{u}rgen},
  title     = {Linear Transformers Are Secretly Fast Weight Programmers},
  booktitle = {Proceedings of the 38th International Conference on Machine Learning},
  series    = {Proceedings of Machine Learning Research},
  volume    = {139},
  pages     = {9355--9366},
  year      = {2021},
  publisher = {PMLR}
}

@article{sommers1987path,
  author  = {Sommers, H.-J.},
  title   = {Path-Integral Approach to {I}sing Spin-Glass Dynamics},
  journal = {Phys. Rev. Lett.},
  volume  = {58},
  pages   = {1268--1271},
  year    = {1987},
  doi     = {10.1103/PhysRevLett.58.1268}
}

@article{sompolinsky1986temporal,
  author  = {Sompolinsky, Haim and Kanter, Ido},
  title   = {Temporal association in asymmetric neural networks},
  journal = {Phys. Rev. Lett.},
  volume  = {57},
  number  = {22},
  pages   = {2861--2864},
  year    = {1986},
  doi     = {10.1103/PhysRevLett.57.2861}
}

@misc{supplementalmaterial,
  note = {See Supplemental Material at [URL will be inserted by publisher] for detailed
          derivations of the generating functional, quenched disorder average,
          saddle-point equations, and mean-field dynamical equations.}
}

@article{wang2010neurophysiological,
  author  = {Wang, Xiao-Jing},
  title   = {Neurophysiological and computational principles of cortical rhythms in cognition},
  journal = {Physiol. Rev.},
  volume  = {90},
  pages   = {1195--1268},
  year    = {2010},
  doi     = {10.1152/physrev.00035.2008}
}

@article{xue2025critical,
  author  = {Xue, Shuyue and Maghrebi, Mohammad and Mias, George I. and Piermarocchi, Carlo},
  title   = {Critical dynamics and cyclic memory retrieval in non-reciprocal {H}opfield networks},
  journal = {SciPost Phys.},
  volume  = {19},
  number  = {4},
  pages   = {100},
  year    = {2025},
  doi     = {10.21468/SciPostPhys.19.4.100}
}

\clearpage
\appendix
\setcounter{secnumdepth}{3} 
\section*{End Matter}

\PRLsection{Summary of the mean-field derivation}
Equation~\eqref{eq:quenched_generating_functional} follows from a
generating-functional calculation~\cite{sommers1987path,coolen2005theory}, sketched here and detailed in the SM~\cite{supplementalmaterial}.
We first promote the local fields to free variables $\theta_{i,t}$
constrained by
$\delta(\theta_{i,t}-h_{i,t})\propto\int d\hat\theta_{i,t}
e^{-\iu\hat\theta_{i,t}(\theta_{i,t}-h_{i,t})}$,
so that the trajectory average is taken at prescribed fields
$\bm\theta$ and the disordered patterns $\bm{\hat\xi}_i$ survive only
through the overlaps
\begin{equation}
    \mu_{\upsilon,t}^a
      = \tfrac{1}{\sqrt N}\sum_i \hat\xi_{i,\upsilon}^a x_{i,t},
    \qquad
    \nu_{\upsilon,t}^a
      = \tfrac{1}{\sqrt N}\sum_i \hat\xi_{i,\upsilon}^a
        \iu\hat\theta_{i,t},
\end{equation}
coupled bilinearly as $\sum_t\bm\nu_t^\top\bm{\hat{A}}\bm\mu_t$ and
imposed, like every order parameter below, by a delta function with their conjugate variables $\iu\hat\mu_{\upsilon,t}^a$,
$\iu\hat\nu_{\upsilon,t}^a$.
The resulting average over $\hat\xi_{i,\upsilon}^a=\pm1$ factorises over
sites, resulting in $\ln 2\cosh(\cdot)$ exponents, that expand to second order in $N^{-1/2}$ into a quadratic in $\iu\hat{\bm\mu},\iu\hat{\bm\nu}$ expressions. 
Integrating out $\iu\hat{\bm\mu},\iu\hat{\bm\nu}$ contributes an exponent
$\tfrac12\ln\lvert\bm Q^{-1}\rvert$, with
$\bm Q = \Bigl(\begin{smallmatrix}
    \bm I_P\otimes\,\bm q & \bm\kappa\\[2pt]
    \bm\kappa^\top    & \bm I_P\otimes\,\bm\rho
\end{smallmatrix}\Bigr)$,
and the conjugates follow from its derivatives.
The saddle point solution  results in the kernels $\bm R$ and $\bm K$ in Eqs.~\eqref{eq:R_saddle}--\eqref{eq:K_saddle} 
in the main text.
Finally, the surviving term $\tfrac12\sum_{t,s}\alpha R_{t,s}
\iu\hat\theta_{i,t}\iu\hat\theta_{i,s}$ is linearised by a
Hubbard--Stratonovich transformation introducing
$\bm z_i\sim\mathcal N(\bm 0,\alpha\bm R)$, after which the integral
over $\iu\hat{\bm\theta}$ collapses $\theta_{i,t}$ onto
$\hmf_{i,t}^{\bm\sigma}$, resulting in a single-spin process.
Because $\bm\chi$ is strictly lower triangular, $\bm\kappa$ is
block triangular in time, so $\bm K$ is strictly retarded and all
kernels propagate forward in time.

\PRLsection{Diagonal exclusion}
The constraint $J_{ii}=0$ removes a self-coupling
$N^{-1}\hat{\bm\xi}_i^\top\hat{\bm A}\hat{\bm\xi}_i\,x_{i,t}$
from the effective field. By the law of large numbers this
concentrates at $\alpha\Tr(\hat{\bm A})/P$ as $N\to\infty$.
Meanwhile, the $n=0$ term in the resolvent expansion
\eqref{eq:K_eigenvalues} gives
$K_{t,t} = P^{-1}\sum_{r=1}^{P}\hat\lambda_r = \Tr(\hat{\bm A})/P$,
since $\bm\chi^n$ for $n\geq 1$ has vanishing diagonal.
The two contributions cancel exactly, so the retarded kernel
in the mean-field equations is strictly causal, $K_{t,s}=0$
for $s\geq t$.

\PRLsection{Numerical setup}
All results in the paper are obtained for $\Delta=0.1$, an intermediate point between fully parallel ($\Delta=1$) and sequential ($\Delta\to 0$) updates.
Mean-field behaviour is calculated for  two settings for the couplings $\bm A^\upsilon$: (i) identical matrices $\bm A^\upsilon = \bm A^0$, thus $\bm{\hat{A}}$ has a finite number of conjugate eigenvalues, and (ii) random orthonormal matrices  $\bm A^\upsilon$, with eigenvalues uniformly distributed on the unit circle. For each case we will compare $2{\times}2$ matrices $\bm A^0$ encoding limit cycle attractors and a $4{\times}4$ matrix  $\bm A^0$ encoding a chaotic attractor.
In the conjugate eigenvalue case, mean-field equations are solved by Monte Carlo sampling of the effective single-spin process with $N=20,000$ spins, and validated against microscopic simulations of $N=500,000$ spins.
Results in Figure~\ref{fig:Fig3} are generated from $N=50,000$ simulations.
All results are averaged over $10$ disorder realizations, with error bars given by the sample standard deviation with the $n-1$ (Bessel) correction.

\PRLsection{Phase diagram of two-pattern cycles at $\alpha=0$}
We consider dynamical attractors constructed by matrices $\bm A = \bm \Omega_\phi$, being $\bm \Omega_\phi$ a $2\times 2$ rotation matrix with eigenvalues $e^{\pm\iu\phi}$. Encoding a single attractor set ($\alpha=0$) , the dynamics converges to the mean field $\bm m_{t+1} = (1-\Delta) \bm m_t + \tfrac{\Delta}{2^M} \sum_{\bm \sigma} \bm \sigma \tanh[\beta \sigma^\top \bm \Omega_\phi \bm m_t]$. This system was studied by \cite{xue2025critical} at the $\Delta\to 0$ limit. The system converges to either one fixed point, four fixed points, or a single limit cycle, where $\phi$ controls the oscillation frequency (radians per update step) as $\omega = \arctan\big[\tfrac{\Delta\beta\sin\phi}{(1-\Delta)+\Delta\beta\cos\phi}\big]$.

We characterize the single-attractor phase structure of the system as a function of rotation angle $\phi$ and inverse temperature $\beta$ by computing three critical lines numerically.
First, we calculate analytically a Hopf bifurcation   by linearizing the overlap dynamics around $\bm m=0$, located at $\beta_c(\phi)  = \Delta^{-1}(-(1-\Delta)\cos\phi   + \sqrt{(1-\Delta)^2\cos^2\!\phi + \Delta(2-\Delta)})
$.
Next, there is a heteroclinic bifurcation at $\beta^*(\phi)$ in which the cycle turns into four fixed points. We locate it using the bisection method for each fixed $\phi$ and finding numerically the value of $\beta$ where the limit cycle disappears. Finally, we calculate a third boundary $\beta_\chi$, characterized by the spectral radius of the response matrix $\bm\chi$ equal to $1$. We locate it numerically, also via the bisection method. After this point, the values of $R_{tt}$ diverge as $t\to\infty$, making the attractor unstable when $\alpha>0$.

\begin{figure}
    \centering
    \includegraphics[width=0.8\linewidth]{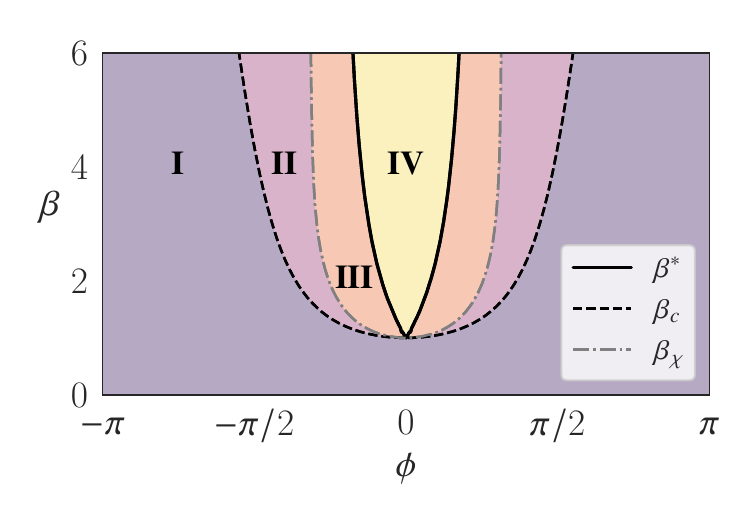}
    \vspace{-0.6cm}
    \caption{Phase diagram of the $\alpha=0$ single-attractor dynamics
    in the $(\phi,\beta)$ plane.
    Three critical lines separate four phases:
    the retrieval onset $\beta_c(\phi)$ (dashed), the cycle--fixed-point
    boundary $\beta^*(\phi)$ (solid), and the noise-stability boundary
    $\beta_\chi$ (dash-dotted) defined by the response function of $\bm\chi$ being larger than one.
    }
    \label{fig:Fig4}
\end{figure}

The three lines partition the $(\phi,\beta)$ plane into four phases
(Figure~\ref{fig:Fig4}).
Region~\textbf{I}: $\beta<\beta_c$, paramagnetic phase with no pattern
retrieval.
Region~\textbf{II}: $\beta_c<\beta<\beta_\chi$, unstable limit-cycle retrieval with $\lim_{t\to\infty} R_{tt} = \infty$, at any load $\alpha>0$.
Region~\textbf{III}: $\beta_\chi<\beta<\beta^*$, stable limit-cycle retrieval with bounded $R_{tt}$.
Region~\textbf{IV}: $\beta>\beta^*$, fixed-point pattern retrieval.

\PRLsection{Calculation of the memory load capacity}
For each inverse temperature $\beta$ and rotation angle $\phi$, we determine the critical capacity $\alpha_c(\beta)$ by bisection on $\alpha$. We initialize the network in a state aligned with the target patterns and evolve the mean-field equations for $\mathcal T$ time steps. The retrieval quality is measured by the root-mean-square overlap $\|\bm m\| = \sqrt{\langle \|\bm m_t\|^2 \rangle_t}$, averaged over the last $K$ time steps to filter transients. We define $\alpha_c$ as the largest $\alpha$ for which $\|\bm m\|$ exceeds a fraction $f$ of its value at $\alpha=0$, with $f=0.5$ throughout. The bisection terminates after at most $10$ iterations or when the bracketing interval falls below $10^{-4}$. For microscopic simulations, $\alpha_c$ is determined independently for each of $10$ random realizations of the disorder, and error bars report the sample standard deviation across trials.

\PRLsection{Calculation of $\bm K, \bm R$ in the general case}
Defining per-eigenvalue matrices $\bm G_\lambda = \sum_{n=0}^\infty \lambda^{n+1}\bm\chi^n$ and $\bm S_\lambda = \sum_{n,m=0}^\infty \lambda^{n+1}(\lambda^\ast)^{m+1}\bm\chi^n\bm q(\bm\chi^\top)^m$, so that
\begin{align}
    \bm K = \frac{1}{P}\sum_{r=1}^P \bm G_{\hat\lambda_r}, \qquad
    \bm R = \frac{1}{P}\sum_{r=1}^P \bm S_{\hat\lambda_r},
\end{align}
one obtains the recursive equations
\begin{align}
    \bm G_\lambda &= \lambda\left(\bm I + \bm\chi\bm G_\lambda\right),
    \label{eq:recursion_G_general}
\\  \bm S_\lambda &= |\lambda|^2\bm q
        + \lambda\bm\chi\bm S_\lambda
        + \lambda^\ast\bm S_\lambda\bm\chi^\top
        - |\lambda|^2\bm\chi\bm S_\lambda\bm\chi^\top,
    \label{eq:recursion_S_general}
\end{align}
which can be solved forward in time due to the strictly lower-triangular (causal) structure of $\bm\chi$. The special cases discussed in the main text follow by substituting the appropriate eigenvalue distributions.

\PRLsection{Chaotic attractor search}
The 4D orthogonal matrix $\bm{A}^\upsilon$ used for the chaotic attractor results  was obtained
via an island-model genetic algorithm at $\alpha=0$.
The search space consists of the $2$ eigenvalue angles $\phi\in[0,\pi]$ and Givens plane angles $\psi_\mu\in[0,2\pi)$ parameterizing $\bm{A}\in\mathrm{SO}(M)$,
together with $\beta$.
The fitness function penalizes the leading Lyapunov exponent $\lambda_1$ falling below a threshold and the spectral radius of $\bm\chi$ exceeding unity, while rewarding positive $\lambda_1$.
Each of the $10$ islands maintains an independent sub-population of 20 individuals evolved with Gaussian mutation with per-island adaptive step size~$\sigma$,
and elitism.
Crossover samples each gene uniformly from an interval extending a fraction~$0.3$ beyond the range spanned by the two parents.
Every 10 generations, the two fittest individuals migrate along a
ring topology. Mutation rates adapt between migrations: $\sigma$ shrinks upon
improvement and grows upon stagnation, balancing exploitation and exploration across islands.
The optimal solution found is
\begin{align}
     \bm{A}^\upsilon = 
     \begin{bsmallmatrix}
      0.72192159 &  0.36388779 &  0.38903112 &  0.44166693 \\
     -0.42269013 &  0.72675866 & -0.35727641 &  0.40682732 \\
     -0.32417986 &  0.41701206 &  0.72090032 & -0.44867704 \\
     -0.44166693 & -0.40682732 &  0.44867704 &  0.66189936
     \end{bsmallmatrix},
\end{align}
at $\beta = 14.7329$, with leading Lyapunov exponents $\lambda_1\approx 0.032$, $\lambda_2\approx -0.001$.

\clearpage




\end{document}


\title{Supplemental Material for: Storing Infinite Dynamical Attractors in Nonreciprocal Associative Neural Networks}
\author{Miguel Aguilera}
\email[Corresponding author: ]{maguilera@bcamath.org}
\affiliation{BCAM -- Basque Center for Applied Mathematics, 48009 Bilbao, Spain}
\affiliation{IKERBASQUE, Basque Foundation for Science, 48009 Bilbao, Spain}
\author{Daniele De Martino}
\affiliation{Biofisika Institute (CSIC, EHU), 48940 Leioa, Spain}
\affiliation{IKERBASQUE, Basque Foundation for Science, 48009 Bilbao, Spain}

\maketitle

Here we provide more details of our derivations and methods.

\section{Dynamical mean-field theory derivation}
\label{sec:dmft_derivation}

We consider a network of $N$ Ising spins $\bm x_t = (x_{1,t},\ldots, x_{N,t})$, $x_{i,t}\in\{\pm1\}$, updated
stochastically in discrete time.  The synaptic matrix has two contributions:
a low-rank structured part built from a set of $M$ target patterns
$\{\bm\xi^a\}_{a=1}^M$ and a disordered part built from $P = \alpha N$ random
patterns,
\begin{align}
    J_{ij}
    =& \frac{1}{N}\sum_{\upsilon=0}^L \sum_{a,b=1}^{M_\upsilon} A_{ab}^\upsilon \xi_{i,\upsilon}^a \xi_{j,\upsilon}^b
    = \frac{1}{N} \bm\xi_i^\top \bm A \bm\xi_j
       + \frac{1}{N} \bm{\hat{\xi}}_i^\top \bm{\hat{A}} \bm{\hat{\xi}}_j, \qquad i\neq j,
    \label{eq:Jij_sm}
\end{align}
where $\bm \xi_{i,\upsilon}\in\{\pm1\}^{M_\upsilon}$, $\bm \xi_i \coloneqq \bm \xi_{i,0}$, $\bm A \coloneqq \bm A^0$ (with $M = M_0$), and $\bm{\hat{\xi}}_i \coloneqq (\bm \xi_{i,1},\ldots, \bm \xi_{i,L})$, $\bm{\hat{A}} \coloneqq \mathrm{diag}(\bm A^1, \ldots, \bm A^L)$, with $L = P / \sum_{\upsilon>0} M_\upsilon$ the number of disorder matrices.

Spin updates are stochastically controlled by independent binary variables $\tau_{i,t} \sim \mathrm{Bernoulli}(\Delta)$: when $\tau_{i,t}=1$ the spin is refreshed via Glauber dynamics; when $\tau_{i,t}=0$ it is copied, $x_{i,t+1}=x_{i,t}$.

For Glauber dynamics, the effective fields $\bm h_t = (h_{1,t},\ldots, h_{N,t})$, take the form
\begin{align}
h_{i,t}
    =& H_{i,t}
       + \sum_{j} \frac{1}{N}
         \big(\bm\xi_{i}^\top \bm A \bm\xi_{j}
           + \bm{\hat{\xi}}_{i}^\top \bm{\hat{A}} \bm{\hat{\xi}}_{j}\big) x_{j,t}
    - \frac{1}{N}
         \big(\bm\xi_{i}^\top \bm A \bm\xi_{i}
           + \bm{\hat{\xi}}_{i}^\top \bm{\hat{A}} \bm{\hat{\xi}}_{i}\big) x_{i,t}
    \nonumber\\
    =& H_{i,t}
       + \sum_{j} \frac{1}{N}
         \big(\bm\xi_{i}^\top \bm A \bm\xi_{j}
           + \bm{\hat{\xi}}_{i}^\top \bm{\hat{A}} \bm{\hat{\xi}}_{j}\big) x_{j,t}
    -\alpha  \hat K x_{i,t}
       + O(N^{-1/2}),
\end{align}
where $\hat K = \Tr(\bm{\hat{A}})/P$ and we used
$\bm\xi_{i,\upsilon}^\top \bm A^\upsilon \bm\xi_{i,\upsilon}
= \Tr(\bm A^\upsilon)
+ \sum_{a\neq b}\tilde{A}_{ab}^\upsilon 
\xi_{i,\upsilon}^a\xi_{i,\upsilon}^b$,
with $\tilde{A}_{ab}^\upsilon = (A_{ab}^\upsilon + A_{ba}^\upsilon)/2$
the symmetric part of $\bm A^\upsilon$
(the antisymmetric part vanishes since
$\xi_{i,\upsilon}^a\xi_{i,\upsilon}^b
= \xi_{i,\upsilon}^b\xi_{i,\upsilon}^a$).
The signal diagonal $\Tr(\bm A)/N = O(1/N)$ vanishes
since $\bm A$ is fixed and finite-dimensional.
The cross terms constitute a sum of $L$ independent bounded
contributions, each of order $O(1)$;
by the central limit theorem the sum is $O(\sqrt{L})$,
giving a correction of order $O(N^{-1/2})$ that vanishes
in the thermodynamic limit.
The only surviving correction is the noise diagonal
$\Tr(\bm{\hat{A}})/N = \alpha \Tr(\bm{\hat{A}})/P$
which is $O(1)$. Later, we will see that this term cancels the $s=t$ contribution of $\bm K$ from the resolvent expansion.

The full transition probability for spin $i$ at time $t$ is
\begin{align}
    p(x_{i,t+1}\mid \tau_{i,t+1},x_{i,t}) = (1-\tau_{i,t+1})\delta_{x_{i,t+1},x_{i,t}} + \tau_{i,t+1}\frac{1}{2}\big(1+x_{i,t+1}\tanh(\beta h_{i,t})\big),
\end{align}
which, for the effective fields above, generates a joint path measure
$p_{\bm h}(\bm x,\bm\tau) = p(\bm x_0)\prod_{t,i} p(\tau_{i,t+1})p(x_{i,t+1}\mid \tau_{i,t+1},x_{i,t}) $.
We denote averages with respect to this measure by
\begin{align}
    \ang{f(\bm x)}_{\bm h} = \sum_{\bm x,\bm\tau} p_{\bm h}(\bm x,\bm\tau) f(\bm x),
\end{align}
following the main text. Note that this notation will allow us to conveniently introduce the effective fields calculated by our mean-field theory later.

We introduce the moment-generating functional
\begin{align}
    Z_{\bm{\hat{\xi}}}(\bm g)
    &= \ang{
    e^{\sum_{i,t} x_{i,t} g_{i,t}} }_{\bm h}.
\end{align}
The derivative of the transition probability with respect to the external field reads
\begin{align}
    \frac{\partial p(x_{i,t+1} \mid \bm x_{t})}{\partial H_{i,t}}
    =& \beta \tau_{i,t+1} \big(x_{i,t+1}-\tanh(\beta h_{i,t})\big)  
    p(x_{i,t+1}\mid\bm x_t),
\end{align}
which gives the general identity
\begin{align}
    \frac{\partial \ang{f(\bm x)}_{\bm h}}{\partial H_{i,s}} =& \beta \ang{f(\bm x) \tau_{i,s+1} (x_{i,s+1} - \tanh(\beta h_{i,s}) )}_{\bm h}.
\end{align}
Using this, functional derivatives of $Z_{\bm{\hat{\xi}}}(\bm g)$ evaluated at $\bm g=\bm 0$ yield standard observables:
\begin{align}
    \frac{\partial Z_{\bm{\hat{\xi}}}(\bm 0)}{\partial g_{i,t}}
    =& \ang{x_{i,t}}_{\bm h},
    &
    \frac{\partial^2 Z_{\bm{\hat{\xi}}}(\bm 0)}{\partial g_{i,t} \partial g_{j,s}}
    =& \ang{x_{i,t} x_{j,s}}_{\bm h},
    \nonumber\\
    \frac{\partial Z_{\bm{\hat{\xi}}}(\bm 0)}{\partial H_{i,t}}
    =& \beta\ang{\tau_{i,t+1} (x_{i,t+1} - \tanh(\beta h_{i,t}))}_{\bm h} = 0,
    &
    \frac{\partial^2 Z_{\bm{\hat{\xi}}}(\bm 0)}{\partial H_{i,t} \partial H_{j,s}}
    =& \beta \frac{\partial}{\partial H_{j,s}}
       \ang{\tau_{i,t+1} (x_{i,t+1} - \tanh(\beta h_{i,t}))}_{\bm h} = 0,
    \nonumber\\
    \frac{\partial^2 Z_{\bm{\hat{\xi}}}(\bm 0)}{\partial g_{i,t} \partial H_{j,s}}
    =& \beta\ang{x_{i,t}\tau_{j,s+1} (x_{j,s+1} - \tanh(\beta h_{j,s}))}_{\bm h}.
\end{align}
These identities relate derivatives of $Z_{\bm{\hat{\xi}}}(\bm g)$ to magnetizations, correlations and response functions.

\subsection{Quenched average over disordered patterns}
\label{sec:quenched_average}

We now compute the quenched average
$\langle\langle Z_{\bm{\hat{\xi}}}(\bm g)\rangle\rangle$
over the disordered patterns $\hat\xi_{i,\upsilon}^a$:
\begin{align}
	\dang{Z_{\bm{\hat{\xi}}}(\bm g)}
    =& \frac{1}{2^{NP}}\sum_{\bm{\hat{\xi}}} 
       \ang{
       \exp\Big[\sum_{i,t} x_{i,t} g_{i,t}\Big]}_{\bm h}.
\end{align}
To proceed, we introduce macroscopic order parameters: magnetizations $\bm m$, fields $\bm \theta$, and auxiliary pattern overlaps $\bm\nu,\bm\mu$; and enforce their definitions using functional delta constraints with conjugate fields. 
In terms of these macroscopic variables we obtain
\begin{align}
	\dang{Z_{\bm{\hat{\xi}}}(\bm g)}
    =& \frac{1}{2^{NP}}
       \sum_{\bm{\hat{\xi}}}
       \int d\bm\eta_1 d\bm\eta_2
       \Bigg\langle
       \exp\Bigg[
           \sum_{i,t} x_{i,t} g_{i,t}
           - \sum_{t,a}\iu\hat m_t^a(\bm g)
             \Big(m_t^a(\bm g)
                 - \frac{1}{N}\sum_i \xi_i^a x_{i,t}\Big)
    \nonumber\\
    &\qquad
           - \sum_{i,t}\iu\hat\theta_{i,t}(\bm g)
             \Big(\theta_{i,t}(\bm g)
                 - H_{i,t}
                 - \sum_j\big(\bm\xi_{i}^\top \bm A \bm\xi_{j}
                              + \bm{\hat{\xi}}_{i}^\top \bm{\hat{A}} \bm{\hat{\xi}}_{j}\big)x_{j,t} + \alpha\hat K x_{i,t}
             \Big)
       \Bigg]\Bigg\rangle_{\bm\theta}
    \nonumber\\
    =& \frac{1}{2^{NP}}
       \sum_{\bm{\hat{\xi}}}
       \int d\bm\eta_1 d\bm\eta_2 d\bm\eta_3
       \Bigg\langle
       \exp\Bigg[
           \sum_{i,t} x_{i,t} g_{i,t}
           - \sum_{t,a}\iu\hat m_t^a(\bm g)
             \Big(m_t^a(\bm g) - \frac{1}{N}\sum_i \xi_i^a  x_{i,t}\Big)
    \nonumber\\
    &\qquad
           - \sum_{i,t}\iu\hat\theta_{i,t}(\bm g)
             \Big(\theta_{i,t}(\bm g)
                 - H_{i,t}
                 - \sum_{a,b}\xi_i^a A_{ab} m_t^b(\bm g) + \alpha  \hat K x_{i,t}\Big)
           + \sum_t \bm\nu_t^\top(\bm g)\bm{\hat{A}} \bm\mu_t(\bm g) 
    \nonumber\\
    &\qquad
           - \sum_{a,t,\upsilon}\iu\hat\nu_{\upsilon,t}^a\Big(\nu_{\upsilon,t}^a
             - \frac{1}{\sqrt N}\sum_i \hat\xi_{i,\upsilon}^a \iu\hat\theta_{i,t}(\bm g)\Big)
           - \sum_{a,t,\upsilon}\iu\hat\mu_{\upsilon,t}^a\Big(\mu_{\upsilon,t}^a
             - \frac{1}{\sqrt N}\sum_i \hat\xi_{i,\upsilon}^a x_{i,t}\Big)
       \Bigg]\Bigg\rangle_{\bm\theta}.
\end{align}
where, for compactness, we denote
\begin{align}
    d\bm\eta_1
    &:= (2\pi)^{-tN} d\bm\theta d\bm{\hat\theta},
    &
    d\bm\eta_2
    &:= (2\pi)^{-tM} d\bm m d\bm{\hat m},
    \nonumber\\
    d\bm\eta_3
    &:= (2\pi)^{-2t\alpha N} d\bm\mu d\bm{\hat\mu} d\bm\nu d\bm{\hat\nu},
\end{align}
and  $\ang{f(\bm x)}_{\bm\theta}$ denotes the average over trajectories $(\bm x,\bm\tau)$ with prescribed fields $h_{i,t} = \theta_{i,t}$.

Next we average over the quenched disorder $\hat\xi_{i,\upsilon}^{a}=\pm1$. The corresponding term factorizes over $i$, and we use a log-cosh expansion to obtain a Gaussian approximation in terms of overlap order parameters:
\begin{align}
     &\prod_{a,\upsilon} \frac{1}{2^{N}}
       \sum_{\bm{\hat{\xi}}}
       \exp\Bigg[
           \sum_{t,i}
           \frac{1}{\sqrt{N}} \hat\xi_{i,\upsilon}^a\Big(
               \iu\hat\mu_{\upsilon,t}^a(\bm g) x_{i,t}
               + \iu\hat\nu_{\upsilon,t}^a(\bm g) \iu\hat\theta_{i,t}
           \Big)
       \Bigg]
    \nonumber\\
    =& \prod_{a,\upsilon}
       \exp\Bigg[
           \sum_i \ln\Bigg(
                2\cosh
               \Big(
                   \sum_t \frac{1}{\sqrt{N}}
                   \big(
                       \iu\hat\mu_{\upsilon,t}^a(\bm g) x_{i,t}
                       + \iu\hat\nu_{\upsilon,t}^a(\bm g) \iu\hat\theta_{i,t}
                   \big)
               \Big)
           \Bigg)
       \Bigg]
    \nonumber\\
    &\simeq \prod_{a,\upsilon}
       \exp\Bigg[
           N\ln 2
           + \frac{1}{2N}\sum_i \sum_{t,s}
               \Big(
                   \iu\hat\mu_{\upsilon,t}^a(\bm g) \iu\hat\mu_{\upsilon,s}^a(\bm g) x_{i,t}x_{i,s}
    \nonumber\\
    &\hspace{7.4em}
                   + 2\iu\hat\mu_{\upsilon,t}^a(\bm g) \iu\hat\nu_{\upsilon,s}^a(\bm g) x_{i,t} \iu\hat\theta_{i,s}
                   + \iu\hat\nu_{\upsilon,t}^a(\bm g) \iu\hat\nu_{\upsilon,s}^a(\bm g) \iu\hat\theta_{i,t} \iu\hat\theta_{i,s}
               \Big)
       \Bigg].
\end{align}

Now, we introduce order parameters for overlap matrices $\bm q, \bm \rho, \bm \chi$
\begin{align}
    q_{t,s}
    =& \frac{1}{N}\sum_i x_{i,t}x_{i,s},
    &
    \rho_{t,s}
    =& \frac{1}{N}\sum_i \iu\hat\theta_{i,t} \iu\hat\theta_{i,s},
    &
    \chi_{t,s}
    =& \frac{1}{N}\sum_i x_{i,t} \iu\hat\theta_{i,s},
\end{align}
together with their conjugates, and collecting all terms into a fourth measure
\begin{equation}
    d\bm\eta_4
    := (2\pi)^{3t^2}
       d\bm q d\bm{\hat q} 
       d\bm\rho d\bm{\hat\rho} 
       d\bm\chi d\bm{\hat\chi},
\end{equation}
we obtain
\begin{align}
    \prod_{a,\upsilon}\cdots
    =& \int d\bm\eta_4
       \exp\Bigg[
           \alpha N\ln 2
           + \frac{1}{2}\sum_{a,\upsilon,t,s}
               \big(
                   \iu\hat\mu_{\upsilon,t}^a \iu\hat\mu_{\upsilon,s}^a q_{t,s}
                   + 2\iu\hat\mu_{\upsilon,t}^a \iu\hat\nu_{\upsilon,s}^a \chi_{t,s}
                   + \iu\hat\nu_{\upsilon,t}^a \iu\hat\nu_{\upsilon,s}^a \rho_{t,s}
               \big)
    \nonumber\\
    &\hspace{3em}
           - \sum_{t,s}\Big(
                 \iu\hat q_{t,s}\Big(q_{t,s} - \frac{1}{N}\sum_i x_{i,t}x_{i,s}\Big)
                 + \iu\hat\rho_{t,s}\Big(\rho_{t,s} - \frac{1}{N}\sum_i \iu\hat\theta_{i,t}\iu\hat\theta_{i,s}\Big)
    \nonumber\\
    &\hspace{3em}
                 + \iu\hat\chi_{t,s}\Big(\chi_{t,s} - \frac{1}{N}\sum_i x_{i,t} \iu\hat\theta_{i,s}\Big)
             \Big)
       \Bigg].
       \label{eq:q_rho_chi_order_parameters_sm}
\end{align}
This step completes the quenched average and introduces the correlation and response order parameters $\bm q$, $\bm \rho$, and $\bm \chi$ that will enter the mean-field theory.

The remaining integrals over $\bm\mu$ and $\bm\nu$ are Gaussian. 
Performing the Gaussian integral over $(\bm\mu,\bm\nu)$ we simplify
\begin{align}
    \int d\bm\mu d\bm\nu e^{\sum_t \bm\nu_t^\top(\bm g) \bm{\hat{A}} \bm\mu_t(\bm g)
       - \sum_{a,t,\upsilon}\big(
           \iu\hat\nu_{\upsilon,t}^a(\bm g) \nu_{\upsilon,t}^a(\bm g)
           + \iu\hat\mu_{\upsilon,t}^a(\bm g) \mu_{\upsilon,t}^a(\bm g)
       \big)}
    =& \exp\Bigg[
        \sum_t \bm{\hat\mu}_t^\top \bm{\hat{A}}^{-1}\bm{\hat\nu}_t
       \Bigg],
\end{align}
and combining with \eqref{eq:q_rho_chi_order_parameters_sm} 
we obtain a quadratic exponent
\begin{align}
    \Lambda
    =& \sum_t \bm{\hat\mu}_t^\top \bm{\hat{A}}^{-1}\bm{\hat\nu}_t
       - \frac{1}{2}
         \bm{\hat\mu}^\top
         (\bm I_P\otimes \bm q) \bm{\hat\mu}
       - \frac{1}{2}
         \bm{\hat\nu}^\top
         (\bm I_P\otimes \bm\rho) \bm{\hat\nu}
       - \bm{\hat\mu}^\top
         (\bm I_P\otimes\bm\chi) \bm{\hat\nu}.
\end{align}

The remaining terms of the exponent are divided in purely macroscopic terms,
\begin{align}
    \Xi
    =& -\sum_{t,a}\iu\hat m_t^a(\bm g)\, m_t^a(\bm g)
       - \sum_{t,s}\Big(
             \iu\hat q_{t,s}(\bm g)\, q_{t,s}(\bm g)
             + \iu\hat\rho_{t,s}(\bm g)\, \rho_{t,s}(\bm g)
             + \iu\hat\chi_{t,s}(\bm g)\, \chi_{t,s}(\bm g)
         \Big),
    \label{eq:Xi_def_sm}
\end{align}
and the microscopic spin degrees of freedom:
\begin{align}
    L =& \sum_{i,t} g_{i,t} x_{i,t}  -\sum_{i,t}
       \iu\hat\theta_{i,t}(\bm g)\Big(
           \theta_{i,t}(\bm g)
           - H_{i,t}
           - \sum_{a,b}\xi_i^a A_{ab} m_t^b(\bm g)
           - \frac{1}{N}\sum_s \iu\hat\chi_{s,t} x_{i,s} - \alpha  \hat K x_{i,t}
       \Big)
       \nonumber \\& + \frac{1}{2N}\sum_{t,s}\iu\hat\rho_{t,s}\sum_i \iu\hat\theta_{i,t} \iu\hat\theta_{i,s}.
\end{align}

Collecting all contributions, the quenched averaged generating functional takes the form
\begin{align}
	\dang{Z_{\bm{\hat{\xi}}}(\bm g)}
    =& \int d\bm\eta_1 d\bm\eta_2 d\bm\eta_4
       \exp\Bigg[
           \Xi
           + \ln \ang{e^{L}}_{\bm\theta}
           + \alpha N\ln 2
           + \ln\Big(
                 \int d\bm\mu d\bm\nu d\bm\eta_3 e^{\Lambda}
             \Big)
       \Bigg],
\end{align}


\subsection{Saddle-point equations and effective single-spin problem}

In the thermodynamic limit $N\to\infty$, the path integral over macroscopic variables is dominated by the saddle point. This yields, at $\bm g = \bm 0$,
\begin{align}
    m_t^a
=& \frac{1}{N}\sum_i
   \xi_i^a\ang{x_{i,t}}_*,
    &
    \iu\hat m_t^a
    =& 0,
    \nonumber\\
    q_{t,s}
    =& \frac{1}{N}\sum_i \ang{x_{i,t}x_{i,s}}_*,
    &
    \rho_{t,s}
    =& \frac{1}{N}\sum_i \ang{\iu\hat\theta_{i,t}\iu\hat\theta_{i,s}}_*
     = 0,
    \nonumber\\
    \chi_{t,s}
    =& \frac{1}{N}\sum_i \ang{x_{i,t} \iu\hat\theta_{i,s}}_*,
\end{align}
where $\langle\cdot\rangle_*$ denotes an average with respect to the effective single-site measure defined by $L$:
\begin{align}
    \ang{f(\bm x,\bm{\hat\theta})}_*
    =& \frac{
        \int d\bm\theta d\bm{\hat\theta} \ang{
        f(\bm x,\bm{\hat\theta}) e^{L}}_{\bm\theta}
       }{
        \int d\bm\theta d\bm{\hat\theta} \ang{ e^{L}}_{\bm\theta}
       }.
\end{align}

The vanishing of the right-hand sides follows from the identities
\begin{align}
    \frac{\partial Z_{\bm{\hat{\xi}}}(\bm 0)}{\partial g_{i,t}}
    =& \ang{x_{i,t}}_*,
    &
    \frac{\partial Z_{\bm{\hat{\xi}}}(\bm 0)}{\partial H_{i,t}}
    =& \beta\ang{\iu\hat\theta_{i,t} }_* = 0,
    \nonumber\\
    \frac{\partial^2 Z_{\bm{\hat{\xi}}}(\bm 0)}{\partial g_{i,t} \partial g_{i,s}}
    =& \ang{x_{i,t} x_{i,s}}_*,
    &
    \frac{\partial^2 Z_{\bm{\hat{\xi}}}(\bm 0)}{\partial H_{i,t} \partial H_{i,s}}
    =& \beta \frac{\partial}{\partial H_{i,s}}
       \ang{\iu\hat\theta_{i,t} \iu\hat\theta_{i,s}}_* = 0,
    \nonumber\\
    \frac{\partial^2 Z_{\bm{\hat{\xi}}}(\bm 0)}{\partial g_{i,t} \partial H_{i,s}}
    =& \beta\ang{x_{i,t} \iu\hat\theta_{i,s} }_*.
\end{align}

The Gaussian integration over the auxiliary fields $\bm{\hat\mu}$ and $\bm{\hat\nu}$ yields
\begin{align}
    \frac{1}{(2\pi)^{Mt}}\int d\bm{\hat\mu} d\bm{\hat\nu} e^{\Lambda}
    =& \frac{1}{(2\pi)^{Mt}}\int d\bm{\hat\mu} d\bm{\hat\nu} 
       \exp\Bigg[
           \sum_t \bm{\hat\mu}_t^\top \bm{\hat{A}}^{-1}\bm{\hat\nu}_t
           - \sum_{a,\upsilon}\Big(
               \tfrac{1}{2}(\bm{\hat\mu}_\upsilon^a)^\top \bm q \bm{\hat\mu}_\upsilon^a
               + \tfrac{1}{2}(\bm{\hat\nu}_\upsilon^a)^\top \bm\rho \bm{\hat\nu}_\upsilon^a
    \nonumber\\
    &\hspace{11.1em}
               + (\bm{\hat\mu}_\upsilon^a)^\top \bm\chi \bm{\hat\nu}_\upsilon^a
           \Big)
       \Bigg]
    \nonumber\\
    =& \frac{1}{(2\pi)^{Mt}}\int d\bm{\hat\mu} d\bm{\hat\nu} 
       \exp\Big[
           -\tfrac{1}{2}\bm{\hat\mu}^\top(\bm I_P\otimes \bm q)\bm{\hat\mu}
           -\tfrac{1}{2}\bm{\hat\nu}^\top(\bm I_P\otimes \bm\rho)\bm{\hat\nu}
    \nonumber\\
    &\hspace{8.6em}
           -\bm{\hat\mu}^\top\big(\bm I_P\otimes\bm\chi - \bm{\hat{A}}^{-1}\otimes\bm I_{\mathcal T}\big)\bm{\hat\nu}
       \Big]
    \nonumber\\
    =& \exp\Big\{\tfrac{1}{2}\ln\abs{\bm Q^{-1}}\Big\}
    \nonumber\\
    =& \exp\Big\{
        -\tfrac{1}{2}\ln\big(
            \abs{\bm I_P\otimes\bm q} 
            \abs{\bm I_P\otimes\bm\rho - \bm\kappa^\top(\bm I_P\otimes\bm q^{-1})\bm\kappa}
        \big)
       \Big\},
\end{align}
where we have introduced the block matrix
\begin{equation}
    \bm Q(\bm q,\bm\rho,\bm\chi)
    =
    \begin{pmatrix}
        \bm I_P\otimes\bm q & \bm\kappa \\
        \bm\kappa^\top      & \bm I_P\otimes\bm\rho
    \end{pmatrix},
    \qquad
    \bm\kappa
    = \bm I_P\otimes\bm\chi - \bm{\hat{A}}^{-1}\otimes\bm I_{\mathcal T}.
\end{equation}
In the third line we used the standard result for a multivariate Gaussian integral, and in the fourth line we used the Schur complement of $\bm I_P\otimes\bm q$ in $\bm Q$.

The conjugate fields associated with the correlation order parameters follow from derivatives of $\ln\abs{\bm Q}$ with respect to the corresponding macroscopic variables. For the $\rho$-sector we obtain
\begin{align}
    \iu\hat\rho_{t,s}
    =& -\frac{1}{2}\frac{\partial}{\partial\rho_{t,s}}\ln\abs{\bm Q}\bigg|_{\bm\rho=\bm 0}
    \nonumber\\
    =& -\frac{1}{2}\sum_{a=0}^{P-1}
       \Big[
           \big(\bm I_P\otimes\bm\rho
              - \bm\kappa^\top(\bm I_P\otimes\bm q^{-1})\bm\kappa
           \big)^{-\top}
       \Big]_{(a\mathcal T+t,a\mathcal T+s)}\bigg|_{\bm\rho=\bm 0}
    \nonumber\\
    =& \frac{1}{2}\sum_{a=0}^{P-1}
       \big[
           \bm\kappa^{-1}(\bm I_P\otimes\bm q)\bm\kappa^{-\top}
       \big]_{(a\mathcal T+t,a\mathcal T+s)}
    \nonumber\\
    =& \frac{1}{2}\sum_{a=0}^{P-1}
       \Big[
           \big(\bm{\hat{A}}^{-1}\otimes\bm I_{\mathcal T} - \bm I_P\otimes\bm\chi\big)^{-1}
           (\bm I_P\otimes\bm q)
           \big(\bm{\hat{A}}^{-\top}\otimes\bm I_{\mathcal T} - \bm I_P\otimes\bm\chi^\top\big)^{-1}
       \Big]_{(a\mathcal T+t,a\mathcal T+s)}
    \nonumber\\
    =& \frac{1}{2}\sum_{a=0}^{P-1}
       \Big[
           \big(\bm I_{MT} - \bm{\hat{A}}\otimes\bm\chi\big)^{-1}
           (\bm{\hat{A}}\bm{\hat{A}}^\top\otimes\bm q)
           \big(\bm I_{MT} - \bm{\hat{A}}^\top\otimes\bm\chi^\top\big)^{-1}
       \Big]_{(a\mathcal T+t,a\mathcal T+s)}.
\end{align}

For the remaining sectors we evaluate the expressions at $\bm\rho=\bm 0$. Using
\begin{align}
    \ln\abs{\bm Q(\bm q,\bm 0,\bm\chi)}
    =& \ln\Big(
           \abs{\bm I_P\otimes\bm q} 
           \abs{\bm\kappa(\bm I_P\otimes\bm q^{-1})\bm\kappa^\top}
       \Big)
    = \ln\abs{\bm\kappa\bm\kappa^\top}
    = 2\ln\abs{\bm\kappa},
\end{align}
we find
\begin{equation}
    \iu\hat q_{t,s}
    = -\frac{1}{2}\frac{\partial}{\partial q_{t,s}}
      \ln\abs{\bm Q(\bm q,\bm 0,\bm\chi)}
    = 0.
\end{equation}
Finally, the conjugate of the mixed order parameter $\chi_{t,s}$ is
\begin{align}
    \iu\hat\chi_{t,s}
    =& -\frac{\partial}{\partial\chi_{t,s}}
      \ln\abs{\bm Q(\bm q,\bm 0,\bm\chi)}
    = -\frac{1}{2}\sum_{a=0}^{P-1}2 \bm\kappa^{-\top}_{(a\mathcal T+t,a\mathcal T+s)}
    \nonumber\\
    =& \sum_{a=0}^{P-1}
       \big(
           \bm{\hat{A}}^{-\top}\otimes\bm I_{\mathcal T}
           - \bm I_P\otimes\bm\chi^\top
       \big)^{-1}_{(a\mathcal T+t,a\mathcal T+s)}.
\end{align}
This makes explicit how the macroscopic correlation and response kernels $(\bm q,\bm\rho,\bm\chi)$ determine their conjugate fields through the inverse of the block matrix $\bm Q$.

Note that  $\iu\hat\chi_{t,s}=0$ for $t>s$.
To see this, reorder the $P\mathcal T\times P\mathcal T$ indices from pattern-first $(a,t)$
to time-first $(t,a)$.
In this ordering, the matrix
$\bm{\hat{A}}^{-\top}\!\otimes\bm I_{\mathcal T} - \bm I_P\otimes\bm\chi^\top$
has $(t,s)$-th $P\times P$ block equal to
$\delta_{ts}\bm{\hat{A}}^{-\top} - \chi_{s,t}\bm I_P$.
Because $\bm\chi$ is strictly lower triangular ($\chi_{s,t}=0$ for
$t\geq s$), the blocks below the diagonal in time vanish, so the full
matrix is block upper triangular.
Its inverse is therefore also block upper triangular, which forces
$\iu\hat\chi_{t,s}=0$ for $t>s$.

Moreover, the retarded kernel $\iu\hat\chi_{t,s}/N$
entering the effective field has a nonvanishing diagonal
$\iu\hat\chi_{t,t}/N = \alpha \Tr(\bm{\hat{A}})/P$,
since expanding the inverse as a Neumann series
$\sum_{n=0}^{\infty}(\bm{\hat{A}}^{n+1}\otimes\bm\chi^n)$
and noting that $[\bm\chi^n]_{tt}=0$ for $n\geq 1$
leaves only the $n=0$ term $[\bm{\hat{A}}]_{aa}$.
This contribution is cancelled exactly by the Onsager correction
$\alpha\Tr(\bm{\hat{A}})/P$ arising from the diagonal exclusion
$J_{ii}=0$, so the diagonal $\iu\hat\chi_{t,t}/N$ always cancels when subtracting the $J_{ii}$ contribution.

\subsection{Integration over conjugate fields and Gaussian effective field}

Substituting $\alpha R_{t,s} = \iu\hat\rho_{s,t}/N$ and $\alpha K_{t,s} = \iu\hat\chi_{s,t}/N$ (note the transposition), we find
\begin{align}
	\dang{Z_{\bm{\hat{\xi}}}(\bm g)}
    =& \int (2\pi)^{-tN}  d\bm{\theta} d\bm{\hat{\theta}}
       \exp\Bigg[
           \Xi
           + \ln \ang{e^{L}}_{\bm\theta}
           + \alpha N\ln 2
           + \ln\Big(
                 \int d\bm\mu d\bm\nu d\bm\eta_3 e^{\Lambda}
             \Big)
       \Bigg],
\end{align}
with
\begin{align}
    L =& \sum_{i,t} g_{i,t} x_{i,t}  -\sum_{i,t}
       \iu\hat\theta_{i,t}(\bm g)\Big(
           \theta_{i,t}(\bm g)
           - H_{i,t}
           - \sum_{a,b}\xi_i^a A_{ab} m_t^b(\bm g)
           - \alpha \sum_{s;\, s<t} K_{t,s} x_{i,s}
       \Big)
       \nonumber \\& + \frac{1}{2}\sum_{t,s} \alpha R_{t,s}\sum_i \iu\hat\theta_{i,t} \iu\hat\theta_{i,s}.
\end{align}
Note that $K_{t,s}=0$ for all $s\geq t$: the terms with $s>t$ vanish by causality, and the $K_{t,t}$ diagonal cancels against the Onsager correction $\hat K$.

The quadratic term in $\hat\theta_{i,t}$ can be linearized by introducing Gaussian auxiliary variables $\bm z$ using the Hubbard–Stratonovich identity:
\begin{align}
\exp\left[
  \frac{1}{2}
  \sum_{t,s}
  \iu\hat\theta_{i,t}
  \alpha R_{t,s}
  \iu\hat\theta_{i,s}
\right]
=& 
\int
\frac{d\bm z}
{\sqrt{(2\pi)^{\mathcal T+1}\det(\alpha\bm R)}}
\exp\left[
  -\frac{1}{2}
  \bm z^\top(\alpha\bm R)^{-1}\bm z
  +\sum_t\iu\hat\theta_{i,t}z_t
\right],
\nonumber\\
    =& \int d\bm{z} p(\bm{z})
       \exp\Big[ \sum_{t}\iu\hat\theta_{i,t} z_{t}
       \Big],
    \\ p(\bm{z}) =& \mathcal N(\bm 0,\alpha\bm R).
\end{align}

Applying this identity and integrating over $\bm{\hat{\theta}}$ gives, spin by spin,
\begin{align}
    \int  (2\pi)^{-\frac{tN}{2}} d\bm{\hat\theta} e^{L}
    = \prod_i \int d\bm z p(\bm z) 
       \prod_t \delta\big(\theta_{i,t}-\hmf_{i,t}\big),
\end{align}
where the effective stochastic field is
\begin{align}
    \hmf_{i,t}
    =& H_{i,t}
       + \sum_{a,b}\xi_i^a A_{ab} m_t^b(\bm g)
       + \alpha\sum_{s;\, s<t} K_{t,s} x_{i,s}
       + z_t,
\end{align}
with $\bm z$ a Gaussian vector of covariance $\alpha \bm R$.

After these manipulations, the quenched average can be written as
\begin{align}
	\dang{Z_{\bm{\hat{\xi}}}(\bm g)}
    =& \int d\bm z p(\bm z) 
       \exp\Bigg[
            \ln\ang{
          \exp\!\bigg[\sum_{i,t} x_{i,t} g_{i,t}\bigg]}_{\bm\hmf}
           + \alpha N\ln 2
           +\tfrac{1}{2}\ln\abs{\bm Q^{-1}}
       \Bigg].
\end{align}

Assuming $H_{i,t}$ is either i.i.d.\ across $i$ or of the form $H_{i,t}=F_t(\bm\xi_i)$, and exploiting self-averaging in the thermodynamic limit, the problem reduces to an effective single-spin description in which $\bm\sigma$ stands for the (random) pattern realization $\bm\xi_i$:
\begin{align}
	\dang{Z_{\bm{\hat{\xi}}}(\bm g)}
    =& \int d\bm z p(\bm z) 
    \frac{1}{2^M}\sum_{\bm\sigma}
       \exp\Bigg[
           \ln\ang{\exp\!\bigg[\sum_{t}x_{t} g_{t}\bigg]}_{\bm\hmf^{\bm\sigma}}
           + \alpha N\ln 2
           + \tfrac{1}{2}\ln\abs{\bm Q^{-1}}
       \Bigg],
\end{align}
with
\begin{align}
    \hmf_t^{\bm\sigma}
    = H_t
      + \sum_{a,b}\sigma^a A_{ab} m_t^b
      + \alpha \sum_{s;\, s<t} K_{t,s} x_{s}
      + z_t.
\end{align}
This is the effective single-spin problem quoted in the main text, Eq.~\eqref{eq:quenched_generating_functional}.

\subsection{Mean-field dynamical equations}

From the effective single-spin problem we can read off the macroscopic dynamical equations, where $\ang{\cdot}_{\bm\hmf^{\bm\sigma}}$ denotes the average over the effective single-spin process with fields $\hmf_t^{\bm\sigma}$. The pattern overlaps obey
\begin{align}
	m_t^a
    =& \frac{1}{2^M}\sum_{\bm\sigma}\sigma^a
       \ang{x_t}_{\bm\hmf^{\bm\sigma}}.
    \label{eq:mean-field-dynamics-patterns_sm}
\end{align}

The two-time correlation function reads
\begin{align}
	q_{t,s}
    =& \delta_{t,s}
       + (1-\delta_{t,s})
         \frac{1}{2^M}\sum_{\bm\sigma}
         \ang{
         \tanh[\beta \hmf_t^{\bm\sigma}]
         \tanh[\beta \hmf_s^{\bm\sigma}]}_{\bm\hmf^{\bm\sigma}},
    \label{eq:mean-field-dynamicx_q_sm}
\end{align}
where the average is over the bivariate Gaussian noise $(z_t,z_s)$ with covariance $\alpha R_{t,s}$.

The response function is
\begin{align}
	\chi_{t,s}
    =& \beta \frac{1}{2^M}\sum_{\bm\sigma}
       \ang{
       x_t\big(
           x_{s+1}
           - \tanh[\beta \hmf_s^{\bm\sigma}]
       \big)}_{\bm\hmf^{\bm\sigma}},
    \qquad t>s,
    \label{eq:mean-field-dynamicx_tilde_q_sm}
\end{align}
which enters the self-consistent determination of the kernels $\bm R$ and $\bm K$.

\section{Response and correlation identities}
\label{sec:response_correlation_identities}

The mean-field equations above involve the response function $\chi_{t,s}$ and the self-correlation $q_{t,s}$, whose structure under partial updates requires detailed bookkeeping of the Bernoulli variables $\tau_{i,t}$. We collect the necessary identities here, where $\ang{\cdot}_{\bm h}$ denotes the microscopic path average.

We derive the explicit form of the response function by specialising the general identities of Section~\ref{sec:dmft_derivation} to $j=i$ and unrolling the stochastic update step by step. Since the variables $\tau_{i,t}$ are independent, we can write, for $t\neq s$,
\begin{align}
    \ang{x_{i,t+1}\tau_{i,s+1} \tanh(\beta h_{i,s})}_{\bm h} =&  \ang{\tau_{i,t+1}\tanh(\beta h_{i,t})\tau_{i,s+1} \tanh(\beta h_{i,s})}_{\bm h} \nonumber \\ & + \ang{(1-\tau_{i,t+1}) x_{i,t} \tau_{i,s+1} \tanh(\beta h_{i,s})}_{\bm h},
    \\ 
    \ang{x_{i,t+1}\tau_{i,s+1} \tanh(\beta h_{i,s})}_{\bm h} =&  \Delta \ang{\tanh^2(\beta h_{i,t})}_{\bm h}, \quad t = s.
\end{align}
We have
\begin{align}
    \ang{(1-\tau_{i,t+1}) x_{i,t} \tau_{i,s+1} \tanh(\beta h_{i,s})}_{\bm h}  = \begin{cases}
            (1-\Delta)\ang{x_{i,t} \tau_{i,s+1} \tanh(\beta h_{i,s})}_{\bm h}, & \text{if $t\neq s$},\\
            0, & \text{if $t=s$}.
           \end{cases}
\end{align}
Then we obtain, for $t>s$,
\begin{align}
    \ang{x_{i,t+1}\tau_{i,s+1} \tanh(\beta h_{i,s})}_{\bm h} = \Delta (1-\Delta)^{t-s} \ang{\tanh^2(\beta h_{i,s})}_{\bm h} + \Delta^2 \sum_{t'=s+1}^t (1-\Delta)^{t-t'} \ang{\tanh(\beta h_{i,t'}) \tanh(\beta h_{i,s})}_{\bm h}.
\end{align}

Similarly, for $t\neq s$,
\begin{align}
    \ang{x_{i,t+1}\tau_{i,s+1} x_{i,s+1}}_{\bm h} =&  \ang{\tau_{i,t+1}\tanh(\beta h_{i,t})\tau_{i,s+1} \tanh(\beta h_{i,s})}_{\bm h} \nonumber \\ & + \ang{(1-\tau_{i,t+1}) x_{i,t} \tau_{i,s+1} x_{i,s+1}}_{\bm h},
    \\ 
    \ang{x_{i,t+1}\tau_{i,s+1} x_{i,t+1}}_{\bm h} =&  \Delta,  \quad t = s.
\end{align}
Also, for $t\neq s$,
\begin{align}
    \ang{x_{i,t+1}\tau_{i,t+1} x_{i,s+1}}_{\bm h} =&  \ang{\tau_{i,t+1}\tanh(\beta h_{i,t})\tau_{i,s+1} \tanh(\beta h_{i,s})}_{\bm h} \nonumber \\ & + \ang{(1-\tau_{i,s+1}) x_{i,t+1} \tau_{i,t+1} x_{i,s}}_{\bm h}.
\end{align}
For $t>s$ we then obtain
\begin{align}
    \ang{x_{i,t+1}\tau_{i,s+1} x_{i,s+1}}_{\bm h} = \Delta (1-\Delta)^{t-s} + \Delta^2 \sum_{t'=s+1}^t (1-\Delta)^{t-t'} \ang{\tanh(\beta h_{i,t'}) \tanh(\beta h_{i,s})}_{\bm h}.
\end{align}

Combining these results yields
\begin{align}
    \ang{x_{i,t+1}\tau_{i,s+1} (x_{i,s+1}-\tanh(\beta h_{i,s}))}_{\bm h} = \Delta (1-\Delta)^{t-s} \ang{1-\tanh^2(\beta h_{i,s})}_{\bm h}.
\end{align}

We also decompose the two-time self-correlation using the update rule. For $t\neq s$,
\begin{align}
    \ang{x_{i,t+1} x_{i,s+1}}_{\bm h} =&  \ang{x_{i,t+1} \tau_{i,s+1}x_{i,s+1}}_{\bm h}  + \ang{(1-\tau_{i,s+1}) x_{i,t+1} x_{i,s}}_{\bm h},
    \\ 
    \ang{x_{i,t+1} x_{i,t+1}}_{\bm h} =&  1.  \quad (t = s)
\end{align}
This gives the recursive relation $q_{t+1,s} = (1-\Delta) q_{t,s}+\ang{\tau_{t+1} x _{t+1} x_s}_{\bm h}$, matching Eq.~\eqref{eq:q_step1} of the main text.

\bibliography{references}